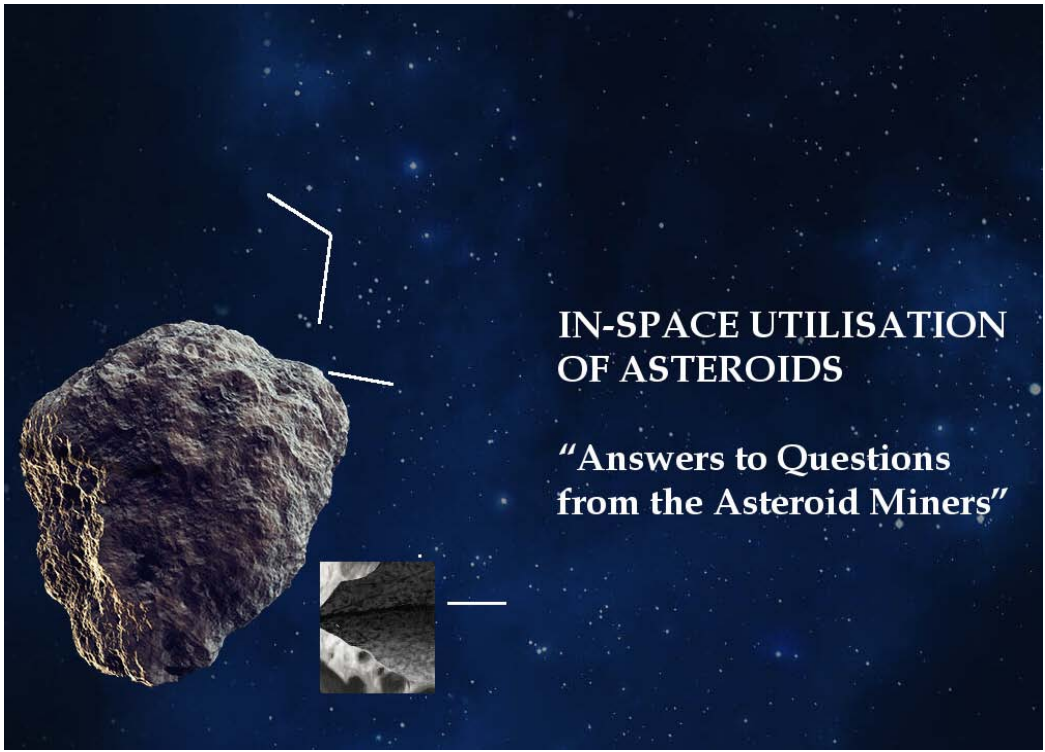

**ASIME 2016 White Paper:**
**IN-SPACE UTILISATION OF ASTEROIDS:**
**"Answers to Questions from the Asteroid Miners"**

**by Amara Graps + 30 Co-Authors**

Outcome from the ASIME 2016: Asteroid Intersections with Mine Engineering, Luxembourg. September 21-22, 2016.

**Date: 15.02.2017**
**Version 1.2**

# AUTHORS

| Amara L. Graps **lead author** | for correspondence: **graps@psi.edu** ) |
| --- | --- |
| Philippe Blondel | Department of Physics, University of Bath, UK |
| Grant Bonin | Deep Space Industries, Inc., USA |
| Daniel Britt | University of Central Florida, Orlando, FL, USA |
| Simone Centuori | Deimos Space S.L.U., Madrid, Spain |
| Marco Delbo | Observatoire de la Côte d'Azur, Nice, France |
| Line Drube | German Aerospace Center, Berlin, Germany. |
| Rene Duffard | Instituto de Astrofisica de Andalucia - CSIC, Granada, Spain |
| Martin Elvis | Harvard-Smithsonian Center for Astrophysics, Cambridge, MA, USA |
| Daniel Faber | Deep Space Industries, Inc., USA |
| Elizabeth Frank | Planetary Resources Inc., USA |
| JL Galache | formerly Minor Planet Center, Harvard-Smithsonian Center for Astrophysics |
| Simon F. Green | School of Physical Sciences, The Open University, Milton Keynes, UK |
| Jan Thimo Grundmann | German Aerospace Center, Koeln, Germany |
| Henry Hsieh | Planetary Sciences Institute, Tucson, AZ USA. |
| Akos Kereszturi | Hungarian Academy of Sciences, Budapest, Hungary. |
| Pauli Laine | University of Jyvaskyla, Finland |
| Anny-Chantal Levasseur-Regourd | Laboratoire Atmosphères, Milieux, Observations Spatiales, Paris, France. |
| Philipp Maier | Max Planck Institute for Extraterrestrial Physics, Garching, Germany. |
| Philip Metzger | University of Central Florida, Orlando, Florida. |
| Patrick Michel | Observatoire de la Côte d'Azur, Nice |
| Migo Mueller | SRON Netherlands Institute for Space Research, Utrecht, Netherlands |
| Thomas Mueller | Max Planck Institute for Extraterrestrial Physics, Garching, Germany |
| Naomi Murdoch | Institut Supérieur de l'Aéronautique et de l'Espace, Toulouse, France |
| Alex Parker | Southwest Research Institute, Boulder, Colorado, USA |
| Petr Pravec | Astronomical Institute of the Academy of Sciences of the Czech Republic, Prague, Czech Republic |
| Vishnu Reddy | LPL, University of Arizona, Tucson, Arizona, USA |
| Joel Sercel | TransAstra Corp. |
| Andy Rivkin | Johns Hopkins University Applied Physics Laboratory, |
| Colin Snodgrass | Open University, Milton Keynes |
| Paolo Tanga | Observatoire de la Côte d'Azur, Nice, France |

Questions were provided by the Companies: Planetary Resources, Deep Space Industries, TransAstra

# Contents

# Introduction

The aim of the Asteroid Science Intersections with In-Space Mine Engineering (ASIME) 2016 conference on September 21-22, 2016 in Luxembourg City was to provide an environment for the detailed discussion of the specific properties of asteroids, with the engineering needs of space missions that utilize asteroids.

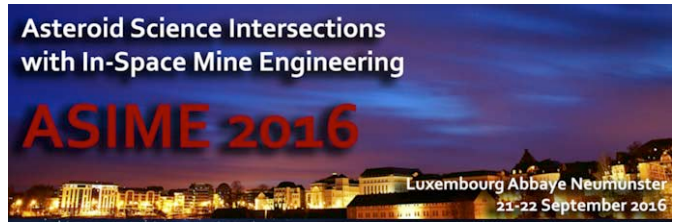

The ASIME 2016 Conference produced a layered record of discussions from the asteroid scientists and the asteroid miners to understand each other's key concerns and to address key scientific questions from the asteroid mining companies: Planetary Resources, Deep Space Industries and TransAstra. These Questions were the focus of the two day conference, were addressed by scientists inside and outside of the ASIME Conference and are the focus of this White Paper.

The Questions from the asteroid mining companies have been sorted into the three asteroid science themes: 1) survey, 2) surface and 3) subsurface and 4) Other. The answers to those Questions have been provided by the scientists with their conference presentations or edited directly into an early open-access collaborative Google document (August 2016-October 2016), or inserted by A. Graps using additional reference materials. During the ASIME 2016 last two-hours, the scientists turned the *Questions from the Asteroid **Miners*** around by presenting their own key concerns: *Questions from the Asteroid **Scientists*** . These answers in this White Paper will point to the Science Knowledge Gaps (**SKG**s) for advancing the asteroid in-space resource utilisation domain.

# Science Knowledge Gaps

1. **More studies are needed to map the classification of meteorites to asteroids.** Presently the best-established link is between ordinary chondrites and S-type **asteroids.** We need more useful published literature about the bulk composition of meteorites to help make more accurate simulants. **We need to understand the meteorite links to C-type asteroids.**

2. **Dedicated NEA discovery and follow-up instrumentation.** The best observability conditions for a given NEA are typically offered around the discovery time (brightest). Need to run observations to characterize NEAs quickly after discovery; best possible with dedicated telescope(s). What is needed: A photometric telescope of a 2-3m class (to reach V ~ 21 with good S/N) available on short notice (for that the observations can be best taken right after discovery). To characterize one NEA, with full IR/vis spectral characterizations, but with 'proxies' or short-cuts to 'each NEO'.

3. **An understanding of granular material dynamics in low-gravity.** Before being sure that we have a robust understanding of the asteroid regolith and to seriously start some systematic material extraction / utilization programs, we must understand how this regolith with its properties responds to the envisage action, i.e. to understand granular material dynamics in low-gravity. Missions like AIM, Hayabusa 2 and OSIRIS-REx can help.

4. **Identifying the available low-delta-v (which are the objects with orbits similar to the Earth) targets are key.** What is needed is a map of low delta-v, low synodic period and low-albedo NEOs as a a first-cut to fine-tune the target possibilities.

5. **Determine if a NEO's dynamically predicted source regions is consistent with its actual physical characterizations.** Knowing the asteroid's source region, and hence, it's orbital family characteristics, can enable a short-cut to characterize the small NEOs of that family which are difficult to measure spectroscopically.

6. **For making useful asteroid regolith simulants, immediate needs are: adequate data on the particle sizing of asteroid regolith and sub-asteroid-regolith surface. How does the asteroid regolith vary with depth?** If the NEOs have structure like comet nucleus 67P, then the NEO regolith is denser then the deep interior.

# I. Asteroid Survey Before Mission Launch

*Here we are in the first phase of a mission that will ultimately land. What we can learn about the relevant global properties before landing and how a landing site might be selected?*

> **1. What are the asteroid properties that asteroid miners need to know? What do we currently know?**

Broadly, what you need to mine an asteroid. (Galache, 2016)

1. Size (radar, IR photometry, stellar occultations, in situ)
2. Composition (Viz+IR spectroscopy, in situ)
3. Mass/Density (Viz+IR spectroscopy, radar if moon(s) present, in situ, Thermal Inertia)
4. Spin characteristics (radar, Viz timed photometry, in situ)
5. Interior Structure (points 1-4, in situ)
6. Surface structure (IR photometry, radar, in situ)
7. Accurate Orbit (Viz+radar astrometry)

We must stress the difference between (Green, 2016):

1) the knowledge that you need for getting to the asteroid and perform the mining,

2) the knowledge that you need that are directly relevant to the mining aspect : i.e. composition, and

Knowledge that we need for choosing / getting to the asteroid (Green, 2016)

- Delta-V is dominant factor for (current) low cost space missions
- As of early November 2016, the number of asteroids known with orbits: 722,669 of which 14,960 are Near Earth Asteroids (NEA) (ref: Minor Planet Center)
- Lowest delta-V targets are subset of NEAs: which have **[SKG 4]**
  - Semi-major axis near 1 A.U.
  - Low eccentricity,
  - Low inclination
- Spin
- Useful tool for planning:
  - NASA Ames Trajectory Browser:
    http://trajbrowser.arc.nasa.gov/index.php

<u>Knowledge of the Asteroid's Composition</u> (Green, 2016; Galache, 2016)

The asteroid's composition is based on *relating* 1) the observational characteristics of the asteroid's <u>taxonomic class (DeMeo et al, 2008)</u> gained from reflectance spectra to 2) meteorite properties.

Meteorite properties. We only have samples from one asteroid (S-type) but we have many meteorites **[SKG 1]**.

Reflectance spectra. We can extend the smaller number of spectral data in visible near-IR wavelengths using spectrophotometry.   (Galache, 2016; Green, 2016, Duffard, 2016)

- Currently determined through ground-based visible, near-IR spectroscopy  (~0.35 - 2.5 µm)  ~2500 asteroids (most only 0.45-1.1 µm)
- Spectrophotometry (3-8 wavebands): ~60,000 asteroids
- Composition may be inferred from albedo, colours, and circularly polarised observations (CPR)
- No dedicated telescopes for NEA spectral classification.  **[SKG 2]**

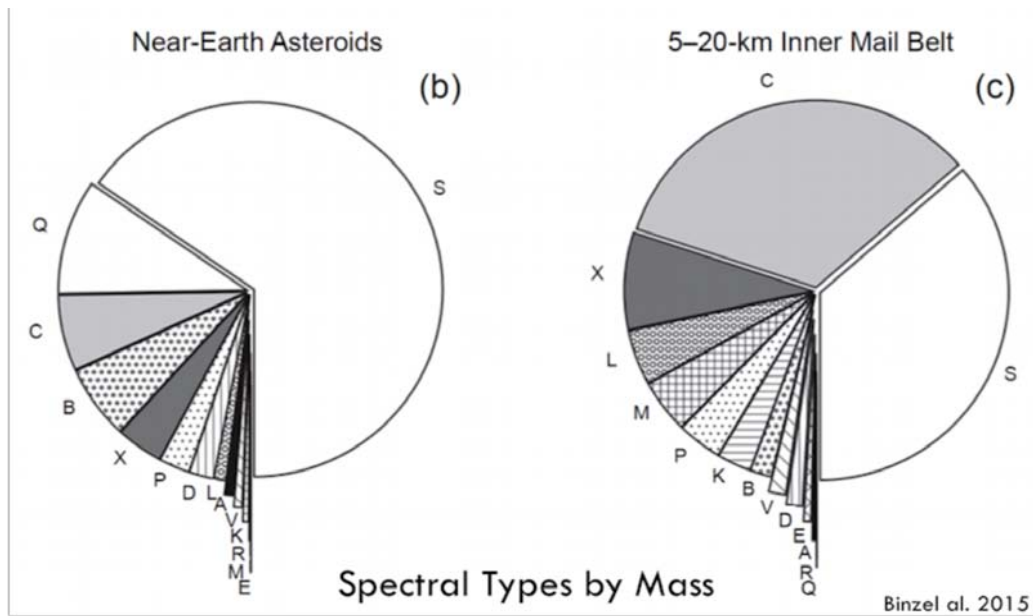

Figure 1 Composition (Viz+IR spectroscopy, in situ). Galache, 2016.

For wavelengths from ~0.4 to ~4 µm, the features diagnostic of composition from reflectance spectra are:
- Continuum shape

- Broad mineral group features
- Subtle narrow mineral features
- Water ice/hydrated minerals (~3.1 μm)
- Organics (3.3 – 3.5 μm)

The caveats for determining the composition with this method.
- Only some components identifiable
- Top few μm of surface only. Our observations only rest on the 'top veneer' of the asteroid, which has been space weathered and has other kinds of processing, which also affect the characteristics of the meteorite samples.
- Uncertain mapping between meteorites and asteroid spectral types **[SKG 1]**

The latter is a point worth repeating: Mapping the classification of meteorites to asteroids is difficult because the **spectral types don't map uniquely to meteorite classes**. **[SKG 1: More studies needed to map the classification of meteorites to asteroids].** See Fig. 2.

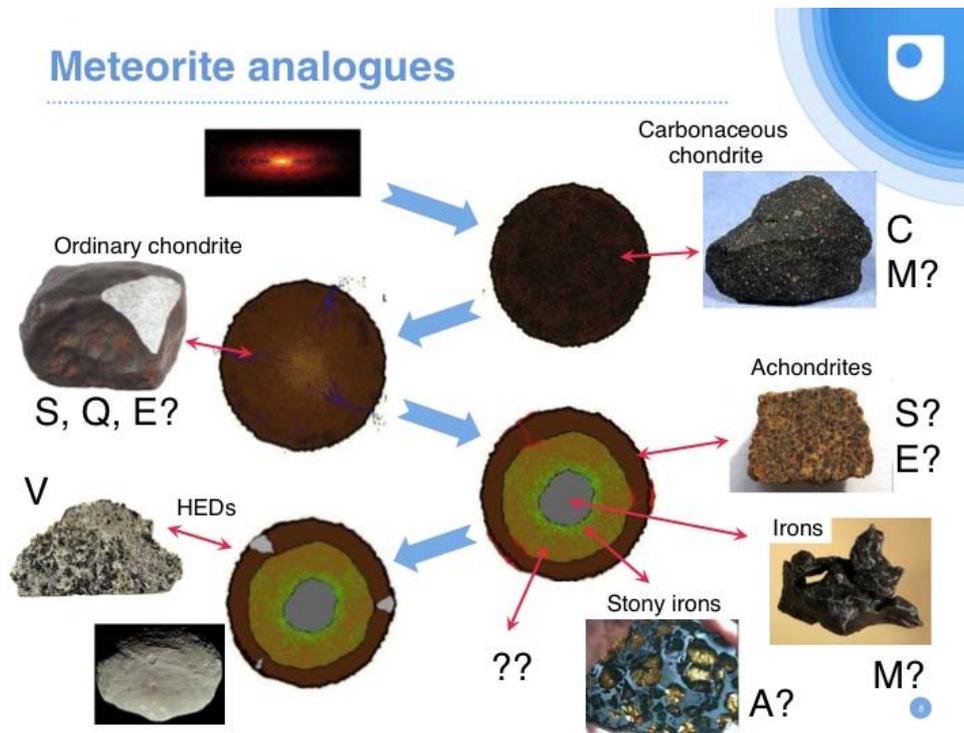

Figure 2 Meteorite Analogues to Asteroid Taxonomic Classes, and lack thereof. From Green, 2016.

**Composition via Thermal Emissivity**

In wavelengths indicative of thermal emission, and only for bright objects or in-situ observations, emissivity features relative to the thermal continuum can provide some composition information.

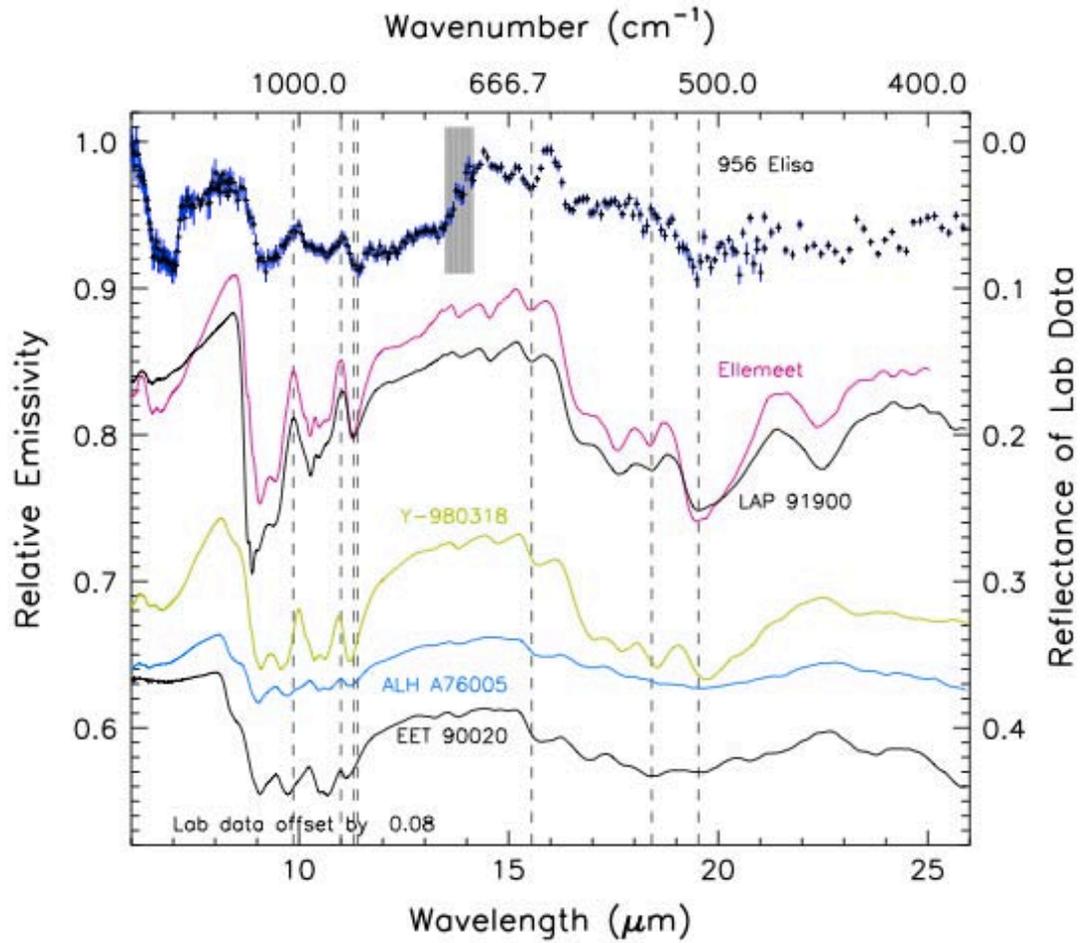

Figure 3. Spitzer spectrum of asteroid 956 Elisa. These features are sensitive to: Silicate mineralogy, crystallline/ amorphous material structure, the regolith particle size. From Green, 2016 and Lim et al, 2011.

**NEO Sizes (radar, IR photometry, stellar occultations, in situ)**
(Galache, 2016)

| Technology | Uncertainty (size) | Uncertainty (mass) | Example (meters) | Num. (/ year) |
|---|---|---|---|---|
| Ground optical survey | 3x | 27x | 30–90 | ~1,800 |
| Space IR | ~20% | 49–73% | 60 ± 12 | 150 |
| Ground radar | 10–20% | 26–73% | 60 ± 12 1,000 ± 100 | Few dozen |
| Ground occultation | 5–15% | 14–52% | 60 ± 6 | ? |

Table gives methods from worst (top row) to best (bottom row).

More details (Green, 2016)
- Discovery magnitude gives diameter to factor ~3. For the follow-up, unless you follow it up *at discovery*, the probability is high that you will 'lose' the asteroid. (Milani, 1999, Galache et al., 2015)
- Albedos (to ~5%) and diameters (to ~10%) from thermal IR fluxes: ~150 000 asteroids from mid-IR surveys
    (WISE/NEOWISE, Warm Spitzer, Akari, IRAS)
  Over 1000 NEOs
  But more than half are from low-precision "Warm" Spitzer data

**NEO Spins** (Green, 2016)

- ~5500 spin periods. from light curves
- YORP effect
    - drives evolution for km-sized NEOs
    - spin up or down in $10^5$ yrs
    - binary creation
    - Navigation issues for fast spinners

See also Fig. illustrating what can we learn from spins in **Answer to Question 7**.

**Composition via Polarimetry / Spectropolarimetery**

Identifying Taxonomic type can also be aided by **polarimetry** (Levasseur-Regourd's and Belskaya, 2016) and Spectropolarimetry (Cellino et al., 2016)

- Polarimetry, can distinguish between low and high albedo asteroidal surfaces, and to estimate taxonomic types through observations above

35 degrees phase angle, quite feasible for NEAs. (Reviews: Cellino et al., 2015, Belskaya et al., 2015)

- Can help identify subsurface properties, e.g., tensile strength, compressive strength, porosity, and of composition and mineralogy, of interest for mining.
- Case Study: Lutetia. Deviation from M-type asteroids, with composition comparable to carbonaceous chondrites and surface heterogeneities, already suspected from polarimetry before Rosetta flew by (Belskaya et al. 2010; Hadamcik et al. 2011).
- Major importance of rendezvous missions that will provide ground-truth on a few objects, and remote studies of many objects. Ground truth for Dawn mission to Vesta (Cellino et al., 2016).
- Spectropolarimetry provides simultaneously spectral reflectance and phase-dependent polarimetric data, and can be used to distinguish between objects belonging to different albedo classes. This technique might soon become the best and fastest technique to achieve a satisfactory physical characterization of asteroid surfaces, the most important limitation being that of the need of large telescopes to obtain useful data for faint objects. (Cellino et al., 2015b)
- More to learn, in a rather near future:
  - Hayabusa 2 and MASCOT to C-type Ryugu,
  - OSIRIS-Rex to B-type Bennu,
  - AIM (with radars) for AIDA mission to Didymos? See Michel, (2016).

What configuration of polarimeter/telescope would one need on a spacecraft to characterize the asteroid? Q&A audio after Levasseur-Regourd's talk (audio Levasseur-Regourd & Belskaya, 2016)?

Answer from Levasseur-Regourd: "If one has a spacecraft in the vicinity of Earth or in deep space, to make a detailed characterization of the asteroid, one needs only a telescope of about 1 m in size, 2 wavelengths, say blue-green, and red, and a simple polarization wheel device to make the measurement at 4 angles for a phase curve. With such a simple configuration, you can get information on that asteroid that goes beyond the simple taxonomic type, for example about an M type."

**Measuring Bulk Density via Thermal IR**

Rozitis et al., (2014) determined the bulk density of small rubble pile asteroid 1950 DA by measurements of its Yarkovsky orbital drift and thermal modelling. The Yarkovsky orbital drift arises on a rotating asteroid with non-zero thermal inertia, and is caused by the delayed thermal emission of absorbed sunlight, which applies a small propulsion force to the asteroid's afternoon side. Thermal-infrared observations can constrain the thermal inertia value, and precise astrometric position measurements conducted over several years can constrain the degree of Yarkovsky orbital drift. Then using the Advanced Thermophysical Model (ATM), in combination with the retrograde radar shape model, archival WISE thermal-infrared data and the asteroid's orbital state, they determined the thermal inertia and bulk density of (29075) 1950 DA. See the next table for the bulk characteristics of the asteroid determined by this method.

### Extended Data Table 2 | Physical properties of (29075) 1950 DA

|  | Property | Value |
|---|---|---|
| Size | Diameter of equivalent volume sphere[8] | 1.30 ± 0.13 km |
|  | Dimensions of dynamically-equivalent and equal-volume ellipsoid (2a, 2b, 2c)[8] | 1.46 × 1.39 × 1.07 km |
| Optical | Absolute magnitude[b] | 16.8 ± 0.2 |
|  | Phase parameter[b] | 0.15 ± 0.10 |
|  | Geometric albedo[8] | 0.20 ± 0.05 |
| Rotation | Rotation period[d] | 2.12160 ± 0.00004 hr |
|  | Obliquity[8] | 168 ± 5 ° |
| Orbit | Semimajor axis[12] | 1.70 AU |
|  | Eccentricity[12] | 0.51 |
|  | Yarkovsky semimajor axis drift[12] | (-2.95 ± 0.57) × 10⁻⁴ AU/Myr (or -44.1 ± 8.5 m yr⁻¹) |
| Surface composition | Spectral type[8] | M |
|  | Thermal inertia* | 24 $^{+20}_{-14}$ J m⁻² K⁻¹ s⁻¹/² 
(36 $^{+30}_{-20}$ J m⁻² K⁻¹ s⁻¹/² at 1 AU) |
|  | Surface roughness* | 50 ± 30 % |
|  | Radar albedo[8] | 0.23 ± 0.05 |
|  | Radar circular polarization ratio[8] | 0.14 ± 0.03 |
| Mass | Bulk density* | 1.7 ± 0.7 g cm⁻³ |
|  | Macro-porosity* | 51 ± 19 % |
|  | Mass* | (2.1 ± 1.1) × 10¹² kg |
| Cohesion | Surface area of negative ambient gravity* | 48 ± 24 % |
|  | Peak negative ambient gravity* | (3 ± 1) × 10⁻⁵ g_E |
|  | Internal cohesive strength* | 64 $^{+12}_{-20}$ Pa |

* Derived in this work.

**2. How can the rate of spectral characterisation of NEOs be increased? It lags far behind discovery rate, especially at smaller sizes (D < 300m).**

This question is aimed towards characterizing the population of potential mining targets for which one would ideally would want characterization of every single object. See also **Answer to Question 8.**

The situation is described in Galache et al., 2015: "The increasing rate of discovery has grown to ~1000/ year as surveys have become more sensitive, by 1 mag every ~7.5 years. However, discoveries of large ($H \leq 22$) NEAs have remained stable at ~365/year over the past decade, at which rate the 2005 US Congressional mandate to find 90% of 140 m NEAs will not be met before 2030 (at least a decade late). Meanwhile, characterization is falling farther behind: Fewer than 10% of NEAs are well characterized in terms of size, rotation periods, and spectral composition, and at the current rates of follow-up it will take about a century to determine them even for the known population. Over 60% of NEAs have an orbital uncertainty parameter, $U \geq 4$, making reacquisition more than a year following discovery difficult; for $H > 22$ this fraction is over 90%."

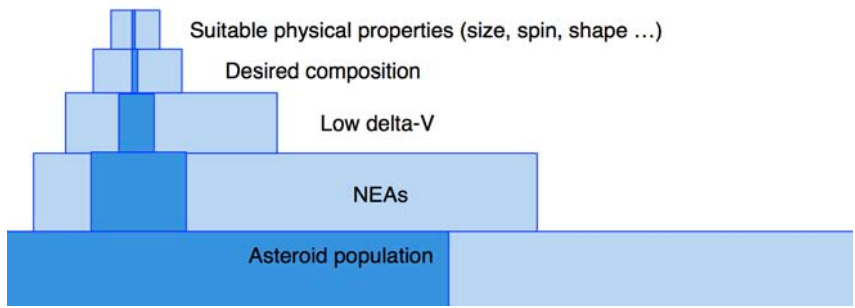

Figure 4. A schematic of the small number of accessible and suitable NEAs for asteroid mining missions. The top level vertical line couldn't be drawn thin enough. (Green, 2016).

A short answer (Green, 2016)
- Some ongoing ground-based observation programmes
    - Need large telescopes – time is competitive, rapid response difficult
    - Dedicated telescope for immediate follow-up is best solution  **[SKG 2]**
    - Exploit new surveys to help

Exploiting **new surveys** can help, but... (Green, 2016)

GAIA  (Operations 2015-2019)
    - Astrometry of all asteroids with V <~20
    - Low-res visible spectra (~150 000 Asteroids)
LSST (operations 2022-2032)
    - >5m asteroid discoveries
    - All-sky survey every few nights in 6 wavebands (0.4-1µm)
Limitations:
    - Spectrophotometric coverage not optimised for solar system objects
    - No IR coverage

Many new targets require follow-up observations
    - More spectroscopy survey targets close to spacecraft
    - New flyby targets
    - Detection of main belt binaries
    - More masses (--> density/ macroporosity) ?
    - Increased family membership statistics

**[SKG 2:** Dedicated telescope needed for discovery and follow-up**].**

From Galache et al., 2015:

"We show that for characterization to keep pace with discovery would require: quick (within days) visible spectroscopy with a dedicated ≥2 m telescope; long-arc (months) astrometry, to be used also for phase curves, with a ≥4 m telescope; and fast-cadence ( < min) light curves obtained rapidly (within days) with a ≥4 m telescope. [...] For the already-known large ($H ≤ 22$) NEAs that tend to return to similar brightness, subsequent-apparition spectroscopy, astrometry, and photometry could be done with 1–2 m telescopes"

To characterize at the discovery rate only optical (0.4-1 micron) spectra/colors can be used as the near-IR sky background is large and variable, limiting observations to V∼<20. The optical sky is dark, so V=22-23 is possible and it makes more sense to observe in the visible wavelengths. This task will still need large (4+ meter, even 8-10 meter) telescopes. Time can be purchased on many of these, although it's not called that.

NEOs require rapid, "target of opportunity" scheduling (within a few hours, or otherwise certainly within the same night). Such scheduling is not in place at most large telescopes. [**SKG 2 again**] Rapid response with a dedicated telescope is needed because NEOs, especially the small NEOs, are usually brightest at the time of discovery; that is the only time when they are bright enough to be observed (see "NEO Sizes" in the previous section). The longer the time period after discovery, when follow-up observations are conducted, the larger one's telescope usually needs to be. That constraint can quickly limit available options.

High quality photometric calibration would also be needed for this approach, but to some extent, this will be provided once the Pan-STARRS1 survey releases its catalog of the northern 3/4ths of the total sky. Note that the discovery rate will grow, possibly greatly, once LSST is observing >2018.

While finding/funding a dedicated telescope is critical for search/discover/follow-up for NEOs [**SKG 2**], a more efficient observational strategy needs to be found rather than just finding a way to apply the brute force approach of doing full IR/vis spectral characterizations of every single NEO, with broadband/narrowband photometry as proxies for spectroscopy.

One observational strategy

Determining the object's spectral slopes would still be useful and can be measured with broadband filters. If possible, narrow-band filters could potentially give you even more information at the expense of not being able to characterize fainter objects. However, the target object's rotation is problematic, when using broadband colors, because one usually observes through one filter at a time, and then subtracts the magnitudes in each filter to derive colors. However, for a rapidly rotating object (which many small asteroids are), the asteroid itself may change magnitude significantly between observations just from rotation.

So for handling rotation, cameras attached to the telescope that obtain observations in multiple filters simultaneously could help with the fast rotating asteroids. These exist but are not widely used. Significant blocks of time on one or more large-ish telescopes (~2m+) with multi-color cameras available could help at least determine spectral slopes for asteroids at a much higher rate than is possible now.

<u>Another observational strategy</u>
A step-by-step approach as in planetary defence (PD) seems useful:
- discovery (present and future PD resources)
- initial orbit determination (here: low delta-v filtering)
- tracking to refine orbit before losing it (same as PD)
- rotation period
- broadband colours
- selected objects: narrowband colours
- selected objects: spectroscopy.

This avoids wasting large-telescope time on objects that are unlikely to be useful. Variability by rotation can be detected by interleaving b/w exposures (clear filter) with colour filter exposures, possibly even compensated while exposure time is short with respect to the rotation period.

## 3. How can we assess the actual number of NEOs of a given spectral type and size?

Three Answers and a discussion.

a) Size distribution (Green, 2016)
- Discovery and re-observation statistics + observation bias models give good indication of NEO population
- D>1 km: ~1000;  D>100 m : 35 000;  D>10 m: ~ 5 million  (Tricario, 2016)

b) Elvis says: ~20,000 NEOs >100m (NEOWISE) and ~20 million NEOs > 20 meters (bolide, infrasound). The shape of the number-size curve in this very steep range is not known and clearly makes a big difference. Efforts are underway with the DECam in Chile to determine this.

c) (Galache, 2016)

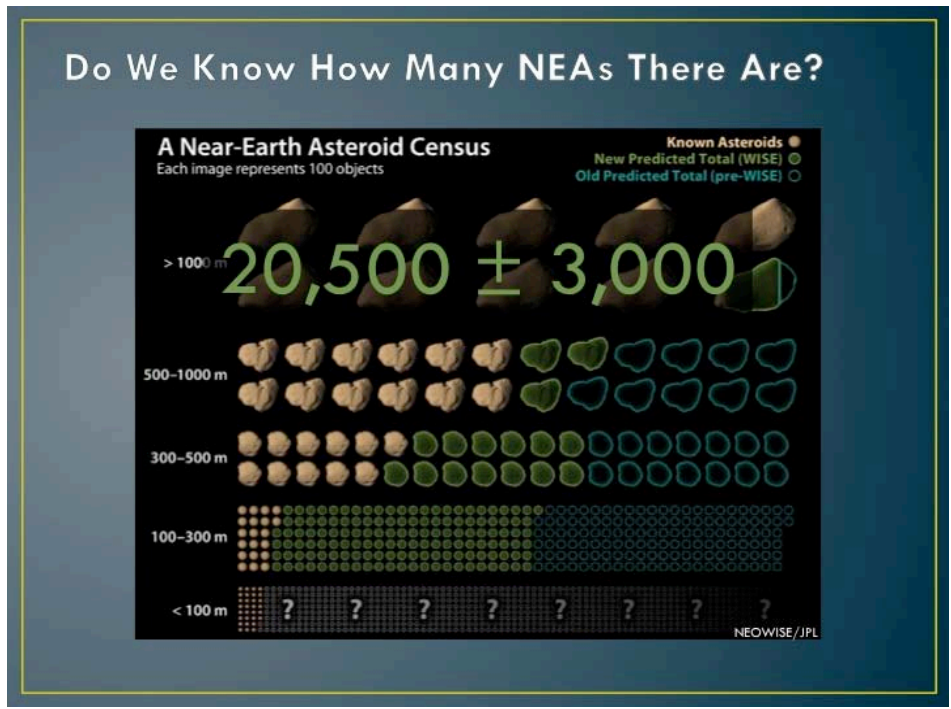

Figure 5. How many NEAs are there? From NEOWISE/JPL and Galache, 2016.

- Spectral type distribution (Green, 2016)
  - Uncertain – relatively small number of NEO spectral types measured
  - Most classes represented but relative numbers will reflect probabilities of ejection and stochastic parent body disruption

It is difficult to make this assessment accurately. The best way is *not* to rely on spectral data (which is hard to acquire), but instead, depend on large surveys such as Gaia and NEOWISE and use their colors/albedos as proxies for composition **See Answer to Question 2**. It is not perfect, but it will get us ahead.

Yes, the spectral types of these small NEOs are still conjecture (Elvis). However (M. Mueller), some information on the taxonomic type can be gleaned from measurements of the albedo (optical reflectivity). In particular, there's a one-to-one correspondence between primitive (and presumably water-rich) composition and low albedo (geometric albedo < ~7.5%). Note that, while hydrated minerals are almost always found on low-albedo asteroids, finding a low-albedo asteroid is not sufficient to guarantee hydrated minerals will be present. For instance, the meteorite Allende is anhydrous for these purposes and is a low albedo carbonaceous chondrite, and the Almahatta Sitta meteorite is low albedo and igneous.

For NEOs, our dominant source of albedo information is from dedicated surveys using Warm Spitzer: ExploreNEOs measured ~500 NEO albedos, NEOsurvey is observing ~500 more (survey ongoing as of September 2016), NEOlegacy will observe ~1,000 more (observations to start in October 2016). Up-to-date info on NEOsurvey can be found at (ref: NeoSurvey)

Caveat for Gaia: It is severely limited in magnitude, as it can't go deeper than V~20 (the real threshold is around 20.5). As a consequence, it can provide spectra for a limited number of NEOs. On the other hand, Gaia's spectral resolution (equivalent to ~30 independent bands) is a good match with the asteroid characterization requirements, but it's limited to the visible range. Diagnostic bands at > 1 micron are out of reach and require complementary observations

An additional hint on the number of NEOs of different sizes can be provided by the dynamical models, most recently the one by Granvik, et al, 2016. With the estimates of relative contributions to the NEO population from different source regions in the main belt, and some trends in appearance of the different main belt groups, it may be possible to make at least some statistical estimates. **[SKG 5]**

> **4. How many NEOs would need to be visited and characterised to create a large enough sample to test current NEO orbital distribution models?**

"None." We would do it from the ground.  (Green, 2016.)

Why?
- remote sensing is adequate
- spacecraft visits of value in confirmation of meteorite-asteroid link

There is no way to get a sample size large enough from spacecraft visits. Ground-based spectral types and a known selection function should be enough to see if NEO orbits and composition are linked.  The few spacecraft visits to asteroids that we do have are still useful though for testing the accuracy of ground-based characterization observations and also for assessing the level of inhomogeneity across asteroid surfaces that could cause ground-based snapshot observations to produce incorrect, or at least incomplete, information.

A more meaningful version of this question might be:

<u>Assuming we can develop models for the likely origins of NEOs based on their dynamical properties (e.g., Greenstreet et al. 2012), can we test these models by seeing if their dynamically predicted source regions correspond with their actual physical characterizations?</u>  **[SKG 5]**

It should be possible with data available right now from WISE/NEOWISE and SDSS (although the current SDSS Moving Object Catalog could stand to be updated).  This is subject to the caveats mentioned later in the answers to the question about whether an asteroid's orbit can be used to infer composition.

**5. Is there any evidence that the shape of an asteroid provides information on its composition?**

Asteroid Shape. Determined by:  (Green, 2016)

- Radar e.g. Greenberg and Margot, 2015.
  - detailed shape
  - limited to close Earth approaches

- Light curves  e.g. Roziti and Green 2014
  - require light curves at multiple apparitions
  - large scale structure only
  - amplitudes give only lower limit to non-sphericity

- Binaries
  ~100 from radar and lightcurves
  - 15% of population

Answer: Not really.
        - The shape may be weak constraint for extreme objects (e.g. fast spinners)

Answer in more detail.

The only place where shape of an asteroid seems to be modified (especially for small NEOs) is by the YORP process, spin up, where the primary would shed mass and form a binary system. I (Reddy) have spectrally characterized a large fraction of YORP binaries in NEO population and I see no evidence for a compositional preference for binary formation. On the other hand, it would be interesting to see if a metallic object can form a YORP binary. I think it comes down to rubble pile vs. monolith and so far we are seeing evidence for rubble piles down to meter sized objects.

Other processes contribute to the shape, such as impacts. The impact response of an asteroid depends on its material composition and porosity. Therefore, the morphology of impact features, that can sometimes contribute to the global shape (e.g. Mathilde, Eros) can tell us something about the composition (a metallic object will show something very different from a carbonaceous one, for instance).

Another clue to the composition by impacts is the secondary ejecta. Secondary (sesquinary) craters are formed by re-impacting slow co-orbital ejecta (Murdoch, 2016).  Such secondary ejecta provides constraints on the ejecta velocity field and knowledge of source crater material properties (Nayak and Asphaug, 2016)

However, the morphology of impact features can only be seen in-situ. Remotely, combined shape, size and spin information allow us to determine whether cohesion is needed to make the asteroid stable in its configuration. It does not say much about the chemical / mineralogical composition, but can tell us whether it needs to be composed of material with some degree of cohesion to be stable with its size, shape and spin (for an assumed density).

**To acquire information on the internal *structure* (not exactly composition) it is useful to evaluate surface structures such as lineaments, crater shapes, crater ejecta, boulder existence / distribution, and mass wasting features**.

They could also provide information on material strength, cohesion, porosity etc. both for the asteroid regolith and interior. The size dependent occurrence of such structures is still poorly known. Such surface morphology information gained right before any mining event could support the planning of the activity (spatial properties, sequential order of steps, especially in-situ analysis). The following two examples indicate the concept.

Largest under-degraded crater
One technique to understand the mechanical properties before any drilling takes place is to note the size of the largest under-degraded crater (Murdoch, 2016, Murdoch et al, 2015). It gives information on the mechanical properties of the body, via estimates of the attenuation of stress energy; see Asphaug, (2008). Weak asteroids may survive collisions that would shatter and disperse monolithic solids as they can can dissipate and absorb energy better.

Grabens or Linear depressions
While asteroids are covered by loose fragmental debris, the surfaces also exhibit grabens or linear depressions (Murdoch, 2016, Murdoch et al, 2015). It is possible that these fractures are evidence of competent rock below the asteroid regolith. It has been suggested that they result from stresses from large impact events, which have refocussed and caused fracture far from the crater (Fujiwara and Asada, 1983; Asphaug et al., 1996), or that they are due to thermal stresses (Dombard and Freed, 2002) and/or body stresses induced by changes in spin. However, faulting can occur even in a granular matrix when it is cohesive relative to the applied stress.

**6. Is there any evidence that the orbit of an asteroid provides information on its composition?**

For main belt asteroids, this is possible due to family affiliation but in a statistical sense. The orbit provides information of the most likely source region.

Although we know that large asteroids are darker in the outer main belt, and brighter in the inner main belt, this "zoning" is not obvious for smaller bodies with NEO sizes  **[SKG 5]**, so it's impossible to assess the composition based solely on the orbit (except, maybe for cometary orbits (see the note below about cometary nuclei).

However, if we know the taxonomic type, then we can estimate the most probable asteroid family that may be at the origin of the object by searching for families close to the most likely source region with the same taxonomic type.

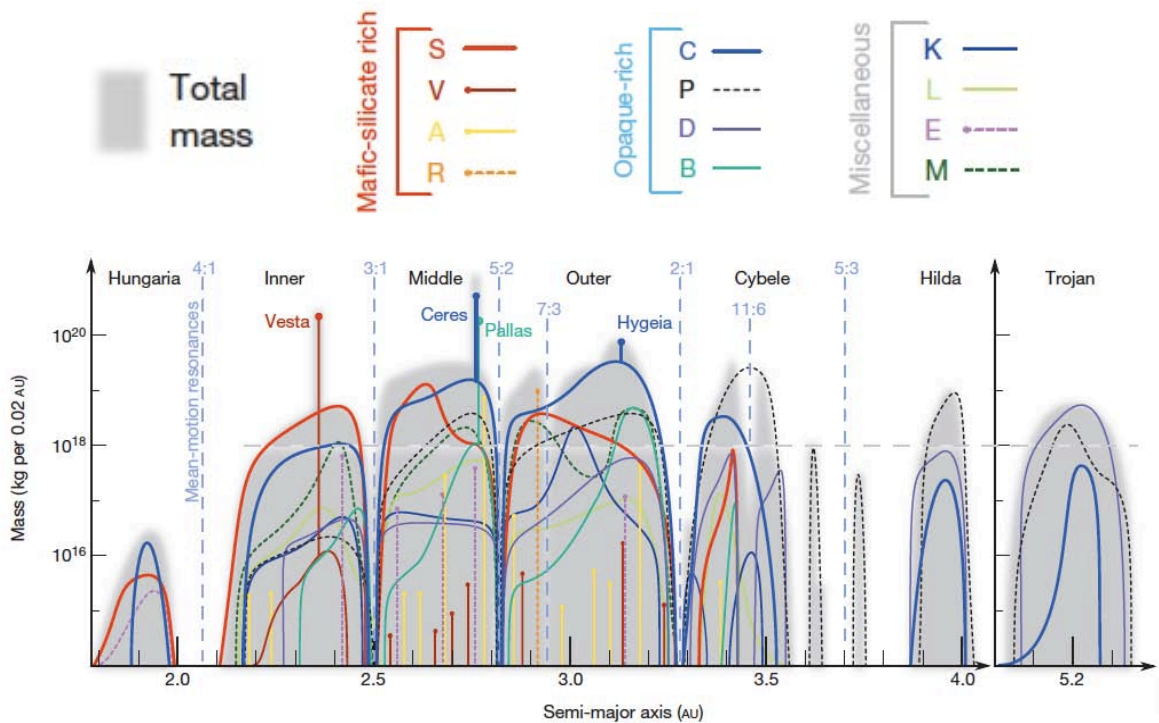

Figure 6 Orbital Distribution. From DeMeo and Carry, 2014 and Green, 2016.

So then for NEOs: First, dynamics can be used to infer the most possible origin in the main belt (Bottke et al., 2002; Greenstreet et al., 2012), and then our knowledge about the composition of main belt asteroids can be used to constrain the composition of the particular NEO in question.  **[SKG 5 again]**

Note, that the above  methodology for NEOs is typically not fully deterministic, chiefly because the dynamics involved is highly chaotic, therefore the information can be gained only in a statistical sense.
- Probability of NEO ejection location (no dynamical memory of original source)

- Greater mixing of types at smaller sizes - influence of Yarkovsky effect?

One prominent exception to the NEO indeterminism: some NEOs are on typically cometary orbits. So in those cases, it is reasonable to assume that their composition resembles that of cometary nuclei.

The following figure provides the current (September 2016) knowledge in numbers (Galache, 2016, Binzel et al, 2015).

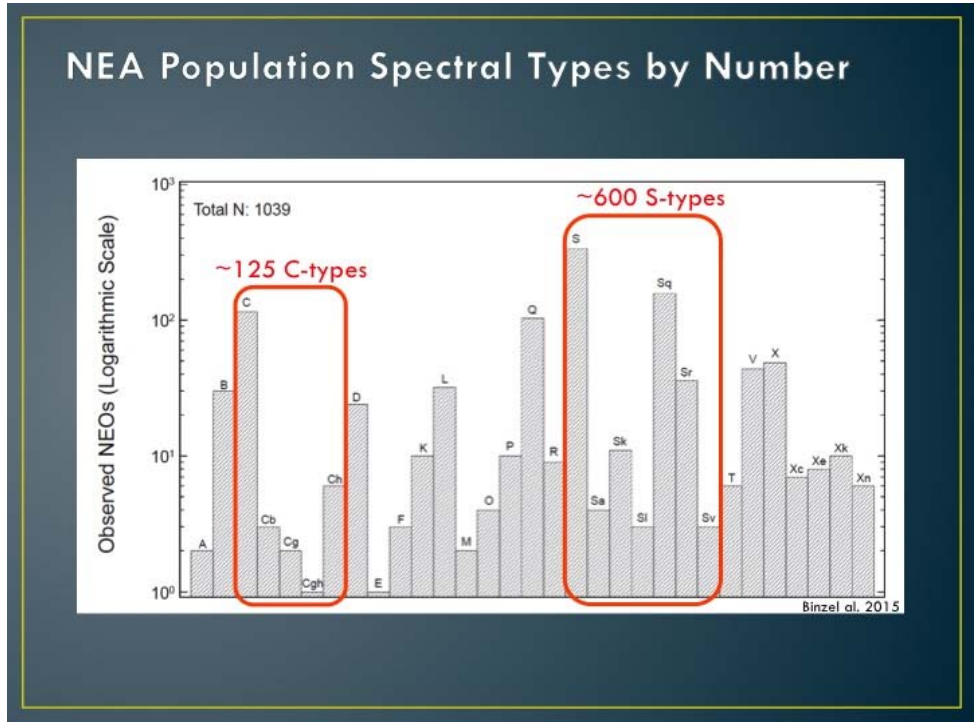

Figure. 7 NEA Population Spectral Types by Number. Figure 5. From Galache, 2016.

Some help from Gaia. One of the big strengths of Gaia will be to provide low-resolution spectra for an amount of asteroids (~350.000) orders of magnitude larger than the currently available sample (a few ~1000s). While it won't be capable to go deeper than magnitude V~20, it will produce (in about 4 years) a new compositional map of the asteroid belt and taxonomy. The challenge will be to couple this amount of information with other surveys in other spectral ranges, and with dynamical properties. This will permit a better characterization of the compositional diversity of the dynamical families (among them, but also internal to each family) and, in the end, a much-improved map of the NEO source regions, both in terms of completeness and detail. We can thus expect that, over a few years, our knowledge of the NEO source will strongly improve.

Gaia is also disclosing the possibility of directly measuring the Yarkovsky acceleration. The approach consists in using the extremely accurate stellar astrometry to calibrate the ground-based astrometry of NEOs, both in new and archive observations. Sensitivity to subtle dynamical effects would thus be much improved. As the magnitude of Yarkovsky depends on the density of the asteroid, such an approach would allow constraining it better. Complementary observations in the thermal IR and thermo physical modeling yielding thermal inertia would strongly enhance the quality of the density determination.

**7. What observations can be made from Earth or Earth orbit that can ascertain the internal structure of an asteroid (rubble pile, fractured shard, etc.)?**

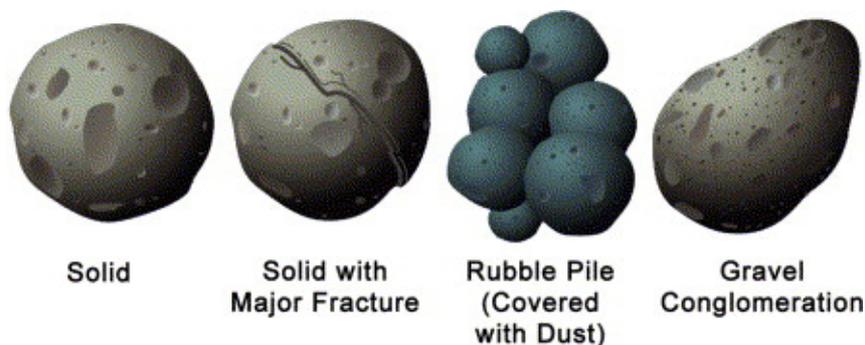

Figure 8. We have never see inside of an asteroid. Which is it? From Walker et al., 2006
and Murdoch, 2016.

Observations about surface and internal structure (Green, 2016; Murdoch, 2016):

- Reflectance spectra or albedo measurements tell us about <u>surface composition</u>.

- Spin rate provides <u>internal strength constraints</u>. Most small asteroids (<100 m) rotate with periods of less than 2h (Pravec et al. 2002; Murdoch, 2016)

The plot in the next figure shows distinct characteristics. (Murdoch, 2016; Scheeres, etal, 2015). No object larger than a kilometer in diameter spins faster than an ~2.4-h period, now known as the gravity spin barrier and that is understood to not be an indicator of cohesionless material by itself. Instead it is simply that at these size scales the gravity effects dominate any strength effect. A number of smaller bodies spin faster; those must experience tensile stress and must have some cohesive strength to hold together. Typically, the observed maximum spins are larger for smaller sizes. Also plotted are spin limits as a function of diameter for levels of asteroid strength.

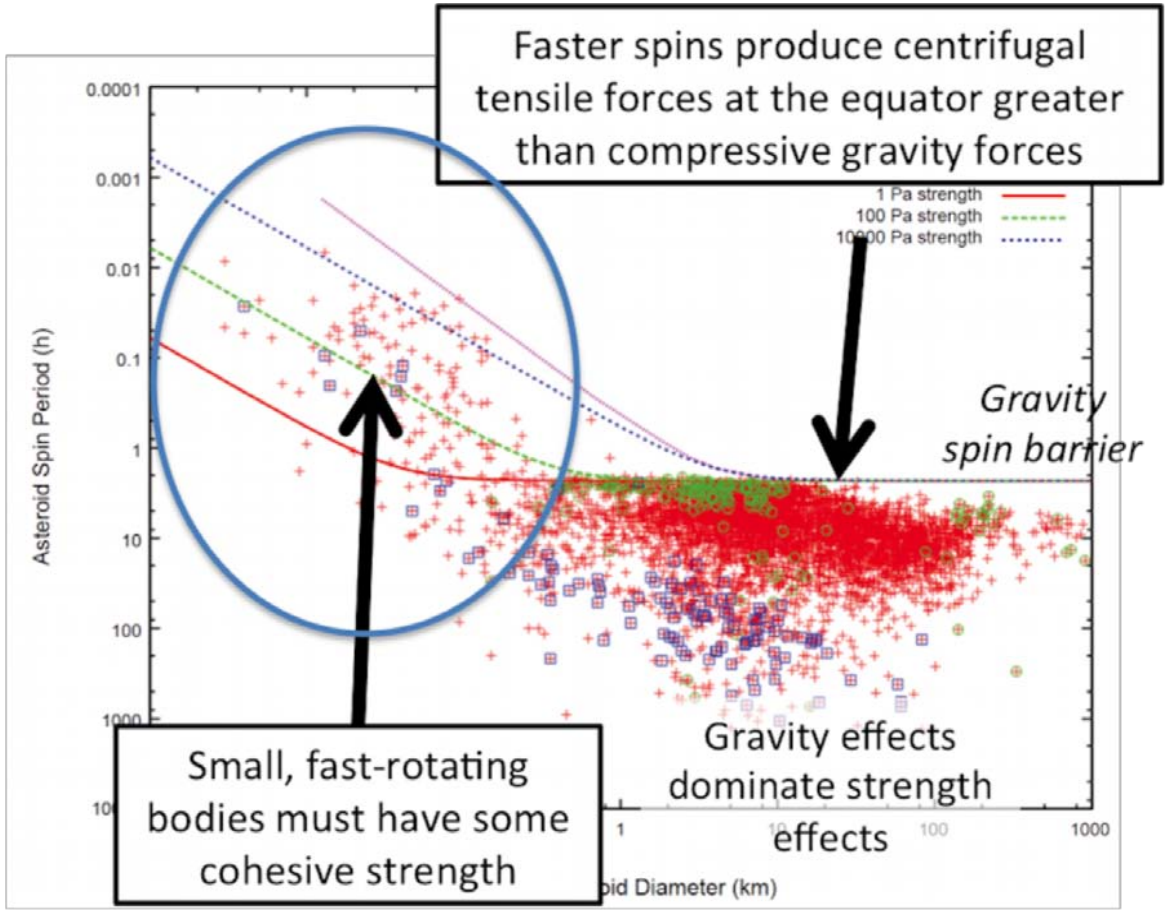

Figure 9. Spin periods versus asteroid diameter in Asteroids IV chapter Scheeres et al, 2015; taken from the database described by Warner et al. 2009 and described further in Murdoch, 2016. There is an observational bias against long periods present – there may exist many small slow rotators we don't know yet (Pravec, 2016).

What is the nature of this cohesion ? (Murdoch, 2016).

| Material | Typical cohesive strength |
|---|---|
| weak soils | 10 kPa |
| weak rocks | 1 MPa |
| moderately strong rocks | 100 MPa |

Asteroid Density

Density measurements are needed to discriminate the composition and internal structure. Densities give a porosity constraint but not structure. To know the internal structure requires both mass estimates (few in number) and volume estimates (large in uncertainties). See Figure below.

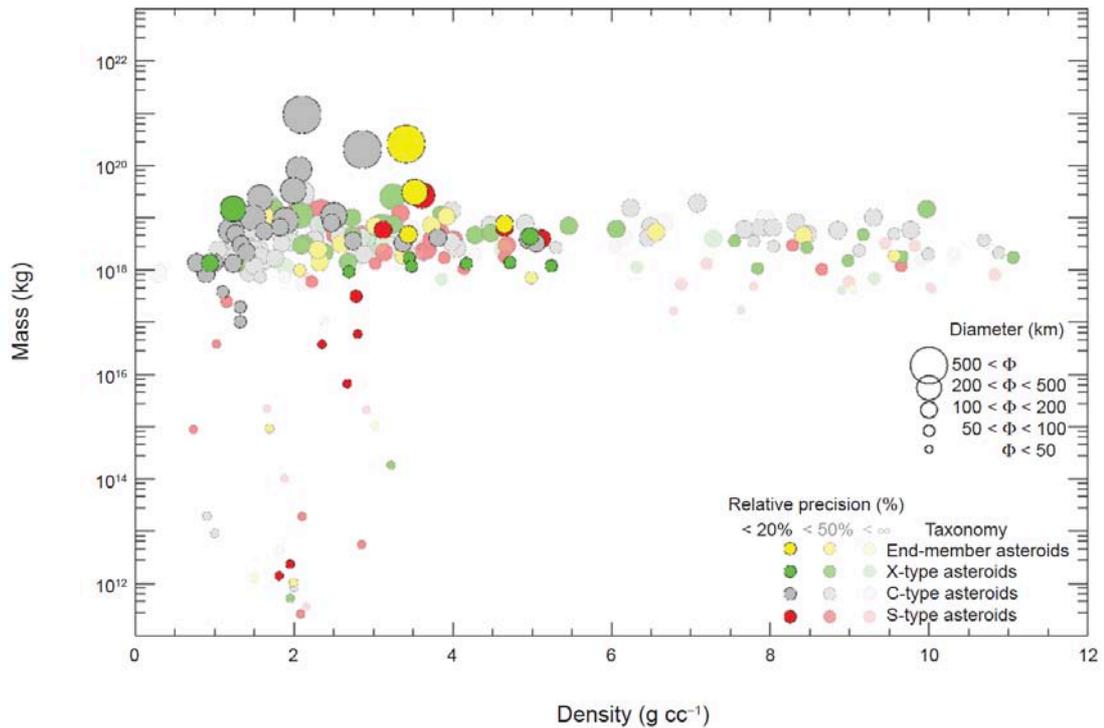

Figure 10. Mass and Density Estimates of Asteroids. Symbol size represents asteroid diameter The colour and contrast of the symbols represent taxonomy and measurement uncertainty, respectively S-type asteroids (red) are on average more dense that C-types (grey) Average S-type: 2.7 g/cm$^3$; Average C-type: 1.3 g/cm$^3$; Average X-type: 1.85 g/cm$^3$ From Carry, (2012) and Scheeres et al., (2015) and Murdoch, (2016).

- Regolith coverage
  - Low thermal inertia for most NEOs implies regolith even on small bodies and fast rotators
  - Van der Waals forces sufficient to retain small particles (<100 Pa) (e.g. Scheeres, etal., 2010)

- Densities
  - From binary periods/separations (radar; light curve eclipse timings + size/shape)
  - Mean porosity from assumed bulk composition

- o Porosities few % to 60% or more → many asteroids are rubble piles
- o Estimating the porosity of an asteroid can tell us about the presence of high density materials (metals) in the interior, or the presence of large voids (or perhaps volatiles)

Asteroid Porosity

C-type asteroids (black) tend to be much more porous than S-type asteroids (red) The largest asteroids have zero macro-porosity. Almost all other asteroids have at least 20% macro-porosity  Average S-type: 31% Average C-type: 50% Average X-type: 87%.

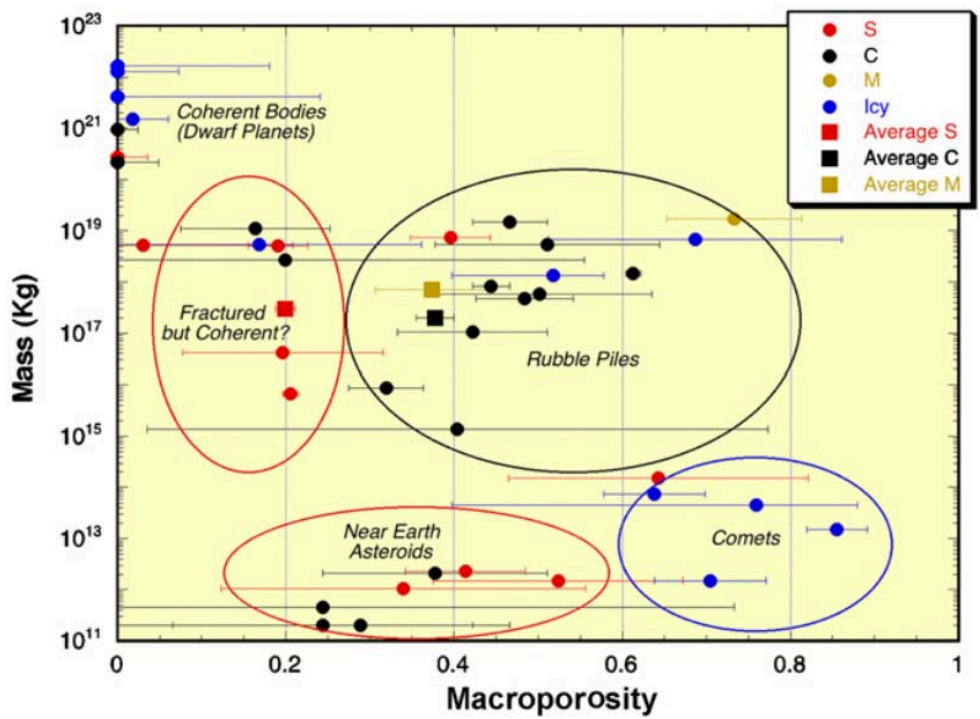

Figure 11. Estimated macro-porosity for a range of small bodies including main belt asteroids, TNOs, near-Earth asteroids, and comets. Asteroid macro-porosity is estimated by subtracting the average porosity of asteroid's meteorite analog from its bulk porosity. Since micro-porosity probably does not seriously affect the structural integrity of small bodies, this is a direct estimate of the large-scale fractures and voids that determine the asteroid's internal structure. Only the largest objects with masses of over 10^20 kg appear to be coherent and have low macroporosity, whereas small bodies have substantial macroporosity. (Consolmagno et al., 2008)

Radar as a Tool of Choice

Boulders can barely be seen on good radar detections. But radar can usually only offer statistical sampling. Distinguishing between rubble piles and other objects on the surface (e.g. hard boulders or large rock formations) still requires close-up observations, due to wavelength/resolution issues. Radar is indeed a tool of choice, either monostatic (using backscatter only) or multistatic (imaging from one direction and receiving from another, as has been done for example with radar imaging of the Moon by the nearby Clementine probe and reception by radio-telescopes on Earth). Optical/visual remote sensing is more easily achieved from Earth orbit, and might provide more information if the objects are close enough or optically reflective enough, but closer views will always be preferred.

Measurements of density via eclipse timing, solar radiation pressure

The mass densities can be derived from careful observations of eclipse timing in binary asteroid systems.  For extremely small NEOs (D<10 m or so), solar radiation pressure can be observed to change their orbit, constraining the area/mass ratio.  Combining this with a size measurement (infrared, radar, …) also provides the mass density.

Surface measurements are inadequate to determine the interior details

Density, global cohesion or angle of friction do not give any details about how the material is arranged *inside* the body (in particular if it's a rubble pile or a fractured shard. And so far, space missions also demonstrated that if we don't have a direct measurement available, we can only infer the internal structure from surface observations with big uncertainties.  As an example, consider the debate regarding Eros, which could equally be a sand pile and a fractured shard, based on the surface images (see Cheng et al, 2007 and references therein). Comet 67P had similar large uncertainties. The only way to ascertain the internal structure is through a dedicated measurement, such as  what is planned in the ESA AIM project (Michel, 2016) with a bistatic low frequency radar.

Multiple asteroid systems can offer clues to the bodies' internal structure.

If we look to the asteroid belt rather than to the NEOs, then close passes between bodies can reveal the mass of the larger body from the deflection of the smaller one. As more objects are discovered, the rate of observable close encounters should increase, though it will always take some luck to be able to find one for an arbitrary object of interest.

The binary nature of an asteroid can be revealed with ground-based photometric or radar observations, and asteroid pairs (in the main belt) are identified by analysis and backward integrations of their heliocentric orbits.]

If we find that the asteroid has or had a satellite (i.e., it's a binary system or an asteroid pair, respectively), the models of asteroid spin fission suggest that it is a rubble pile. Certainly at least there was a negligible cohesion of (near-) surface material from which the satellite formed. Moreover, the equatorial ridge observed on about half of asteroid binary primaries suggests a migration of loose regolith under the tidal forces from the secondary. <u>Note: Tidal forces from the satellite could potentially make the rubble pile interior more unstable and therefore more dangerous for mission landing</u> (comment from Pravec in Q&A audio after Murdoch's talk).

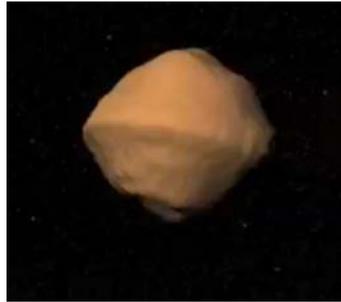

Figure 12. Radar observations of 1999 KW4. Ostro et al, 2006; Pravec, 2006.

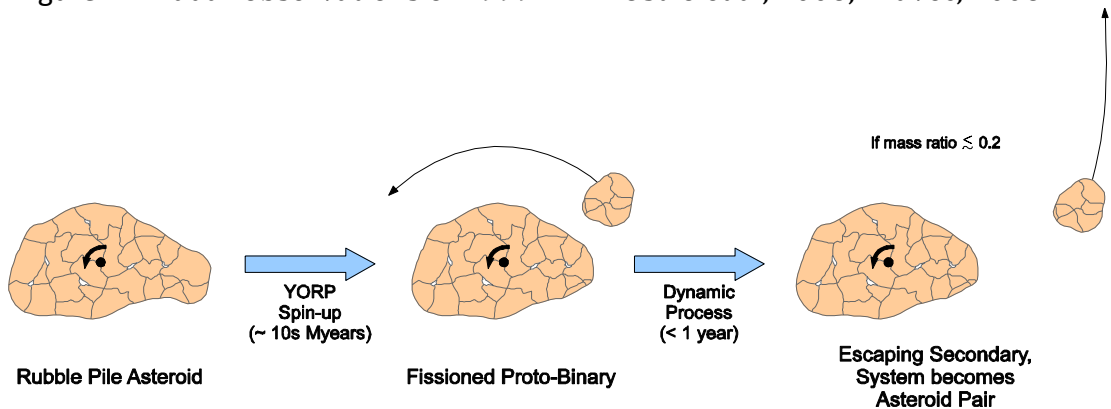

Figure 13. Formation of asteroid pairs by rotational fission. Pravec et al. 2010, Pravec, 2016.

**8. What highest value telescopic composition/characterisation studies are not being pursued for lack of funding or perceived low priority from space agencies? See also Question 2.**

<u>A short answer. Green, 2016.</u>

- Priority for mitigation funding is still on making NEO discoveries
  - Spectroscopic follow-up requires dedicated telescope **[SKG 2 again]**
  - Observations made at discovery maximises S/N
  - Multi-band imaging for fainter targets

- Competition should *not* be with astronomical facilities
  - One-at-a-time targets are hard to compete with multi-object programmes
  - Light curves require extended observations
  - Objectives are not (primarily) science driven

- Large surveys can help
  - Astronomical (non-asteroid) surveys often limited value for NEAs
        Limiting magnitudes to bright;  non-optimal observing strategy
  - Expect some improvement from LSST and Gaia

More detailed answer.

<u>Ground-based facilities</u>

Currently, for ground-based instruments the discovery of new NEOs has larger priority over dynamical or physical characterization. It is clearly difficult to obtain telescope time at large facilities for performing time-demanding surveys (for instance, by spectroscopy) of a large number of targets.  Many "new" or known NEOs are not followed up at all (some of them- actually "many" for small sizes ~140m - can even be "lost" as too few astrometric observations follow discoveries). The situation is not going to improve with the advent of LSST and an increasing rate of discoveries - simply the fraction of characterized NEOs will decrease. <u>To solve this problem, dedicated telescopes are probably the most efficient approach.</u>  .
**[SKG 2 again]**

<u>Ground-based facilities</u>: Dedicated, Robotic Telescopes

Dedicated (possibly, fully robotized) facilities on the ground could be very performing on several aspects (light curves for shape determination, visible-NIR low-resolution spectra, astrometry).
<u>Ground-based facilities</u>: VNIR,  Visual

Some suggested that a large scale VNIR survey of the NEO population would be helpful. Something like <u>1 year dedicated survey</u> on NASA IRTF or UKIRT. If we can produce a SMASS-like survey for small NEOs in the NIR that would be helpful for answering a number of questions here.  Others disagreed by pointing out that NIR can't go fainter than V~20. Because most NEOs are now discovered at V~20-21 and then fade,  NIR is (for now) impractical. Instead it was suggested that we have to use Visual band (0.4-1.0 micron) only. It will depend on what exactly the required rate of false positives/false negatives from the mining side is. Visible wavelengths (the "0.7-μm band") could work if you don't care about false negatives (though there are details to that as well).

Ground-based facilities: Radar

The best method against losing asteroids, particularly relatively small and faint ones near Earth which quickly fade out of telescopic view again, is immediate radar observation on the discovery conjunction. This gives an orbit accuracy comparable to a visual telescope recovery at the next conjunction. But this requires facilities like Goldstone or Arecibo in size, which would have to operate in close contact responding to the visual wide-field surveys' output - very expensive

Ground-based facilities: Multiple filters (colors), instead of spectra and Network of small telescopes

A large-scale color characterization survey for NEOs capable of rapid response to new discoveries and using a camera capable of observing in multiple filters simultaneously could be pretty useful.  Colors would not be as good as spectra, but can at least give you spectral slopes, and simultaneous multi-filter observations will remove uncertainties due to rotational brightness variations. As mentioned above, rapid response will mean that a follow-up telescope may not need to be as large as would otherwise be needed since NEOs, especially small ones, are typically brightest at the time of discovery, and fade quickly as their distance from Earth increases.  A network of such telescopes (even just 2 or 3) would be even more useful to allow follow-up of a larger number of NEO discoveries than would be possible from just a single telescope from which not all newly discovered NEOs will be observable for various reasons.

Ground-based facilities: Harnessing Underused Small Telescopes

There are a number of 1-meter or larger telescopes around that were in scientific observatories but became obsolete for 'big science' or were replaced by more advanced or larger ones. A survey of such under-used, mothballed or become-museum telescopes would be useful to see whether they could be converted for some type of spectroscopy. All of the early Planetary Defence surveys were based on decommissioned light buckets, but in that case, they were wide-angle (in astronomers' terms) designs e.g. made for satellite tracking. Now the more typical astronomy telescopes would be useful which have a large aperture but also small field of view because this better suppresses background light from the atmosphere or unresolved objects. Also, top class amateur telescopes now have passed the 1 m aperture mark.

Dedicated Space-Based Facility/ies

Telescopes in space can have the best observing conditions and access wavelengths that are precluded to ground-based instruments. A study not pursued for lack of funding is the investigation of satellites constellation dedicated to NEOs observation from space. Having such system in space would be surely more expensive and more risky than just relying on a ground-based network, but it would also provide the consistent advantages of being free from the limitations imposed by weather conditions, by the day/night cycles and by the scattered light into the atmosphere. Some food for thoughts: the system could be deployed in a single dawn-dusk SSO, observing in the anti-sun direction perpendicular to the polar plane, to exploit good and stable illumination conditions and to have similar dynamical perturbations for all the satellites (that would make the constellation more stable). The scanning strategy would have to be defined accordingly to a trade-off between tracking and surveillance and it would lead to the selection of the minimum number of satellites. The constellation could be increased along time, but it seems reasonable to have something already functional with 4 spacecraft, thus within a single launch with a heavy carrier. Complexity would be surely high, but the constellation would also have an operational flexibility impossible to be achieved with ground-based systems.

Space-based (single) telescope example: AsteroidFinder (at DLR)

AsteroidFinder was an advanced design, ~150 kg class spacecraft, using an electron-multiplied (EM) CCD and a 25-cm-class telescope. The design was carried through to Phase B (and a B2 and B+) but then discontinued. If flown, the telescope would have pointed towards the Sun to discover and track asteroids at solar elongations >30°. The scientific purpose was to discover more asteroids Interior to Earth's Orbit (IEO) with the goal to at least double the known population, i.e. find another 10 or so in 1 year. It would also have caught hundreds of Aten NEAs >100 m, and a lot of Apollos. Plan B was a conventional CCD with image stabilization. Using a dawn-dusk Sun-synchronous (DDSSO) orbit sped up the discovery rate and improved tracking of discovered or known objects. Summary, such a design could be achieved by one agile <200 kg spacecraft. One could implement two or three, based on the same design with different sunshade shapes.

> ## 9. What is the minimum suite of instruments that could be placed on a prospecting mission to study the cohesive and mechanical strengths of asteroid surfaces (to inform mining strategies)?

Short answer (Green, 2016)

Very little on this can be determined from orbit. Some inferences on regolith and body strength from imaging of local topography. You need to be close to the surface to determine these cohesive and mechanical characteristics, and then you are already there. Knowledge of this ideally requires interaction with the surface. More on that below.

<u>Interacting with the asteroid regolith</u>: How to study its cohesive strength and mechanical properties

Any mission that interacts with the surface will tell us something about its response and therefore its mechanical properties. This can be done through the observation of how a lander bounces or the outcome of an impact, like in the AIDA project with ESA and NASA, and Hayabusa 2 by JAXA, or the response to a sampling tool (Hayabusa 2, OSIRIS-REx by NASA), or a seismic experiment, or an instrument to measure the local surface strength. We have very little understanding regarding the mechanical properties of asteroids, and how they respond to an external action in a low-g environment **[SKG 3].** The mentioned missions are all unavoidable steps towards a better understanding, before asteroid mining can become robust and a reality. These presentations discussed interacting with the regolith in great detail: (Biele et al., 2016; Murdoch, 2016; Michel et al., 2016)

<u>Lessons Learned from the Lunar Missions</u>

In the Surveyor, Apollo, and Lunakhod lunar missions, cohesion and internal friction were inferred from a wide variety of mechanical interactions with the regolith, including: boot impressions, rover tracks, boulder tracks (from boulders naturally rolling downhill, e.g. ref: Apollo 12), terrain slopes, cone penetration resistance, shear cone resistance, scoop digging resistance, manually digging trenches, collapsing the dug trenches, lander footpad sinkage, rocket plume interaction, and many more. Similar methods can be used on asteroids, so basically any mechanical interaction with the regolith will provide information.

<u>How does the asteroid regolith vary with depth?</u>

There's significant uncertainty about how the regolith varies with depth **[SKG 6].**

Since the asteroid miners will be targeting undifferentiated small porous bodies, comet 67 P has lessons. "We found that the density of the subsurface layer is much higher than the density of the deep interior .. " (Levasseur-Regourd, at the Q&A audio after Delbo's talk)

Does electrostatic winnowing produce a "desert pavement" of coarser material on the surface? Is it dustier beneath the surface? Do fines formed by thermal cracking sink into the regolith only a short distance, forming a fine layer of small thickness? The mechanical strength of fine layers will be very different than coarse layers.

Since mining will presumably penetrate the surface deeper than these possibly variations in the regolith, it seems that an important test would be to penetrate or dig deeply into the regolith. This might occur on the OSIRIS-REx mission, because backscatter of regolith from the TAGSAM regolith sampler might disrupt deeply enough to look for variations in the regolith. A prospecting mission could similarly inject gas to disrupt to a desired depth.

Another way to test regolith mechanical variation with depth is to perform a "cone penetration test.": A small spacecraft may not have enough mass and inertia to penetrate to any significant depth, but tests by Honeybee Robotics have shown that the cone can be either percussing or inject gas from its tip (or both) to dramatically reduce penetration resistance while still providing a good correlation between measured resistance and regolith properties. Thus, a small spacecraft might be able to perform a penetration test even in the low gravity of an asteroid.

> **10. What is the minimum suite of instruments that could be placed on a prospecting mission to study the regolith properties of asteroid surfaces (to inform mining and in situ risk mitigation strategies)?**

Short Answer: Green, (2016)

Important for determining if one should land or put instruments on the surface. Imaging systems and radar are useful. Few mm for imaging systems. Radar will help you understand on the order meter. Once on the surface, there are different needs. See the suggested in-situ instruments listed in Section II Asteroid Surface Environment.

- Camera + Thermal imager/spectrometer
  - Distinction between solid rock porous rock/regolith and dominant regolith particle size, inferred from thermal inertia (Gundlach and Blum, 2013)
  - Requires at least 2 wavebands and preferably multiple phases to remove inertia/roughness degeneracy
  - Trend to larger grain size for smaller asteroids
  - Spectrometer (for composition) lower priority

- Radar
    - Some inference of subsurface/interior structure
    - Trade-off with depth and resolution
    - Complex observation strategy, better with surface element and bistatic radar

More Detailed Answer.

The thermal imager needed would cover at least two wavelength bands between 5 and 25 micron.  A low-res spectrometer would also work..  Both OSIRIS-REx and Hayabusa 2 carry such instruments.  The thermal inertia (essentially a proxy for thermal conductivity) of regolith is drastically lower than that of bare rock and decreases further as the typical grain size decreases (fine regolith: extremely low thermal inertia; coarse regolith: moderate thermal inertia; bare rock: high thermal inertia).  The measurement principle is to measure the surface temperature (hence the need for two or more wavelengths) at a number of local times. See **Answer to Question 21** for more discussion.

<u>One clue about the asteroid interior from an orbital distance</u>

If the largest boulders found at the surface are comparable in size to the asteroid itself : they may have been produced during a catastrophic disruption event, and the asteroid is more likely to have a rubble pile structure. See more clues about the asteroid interior from surface features described in the **Answer to Question 5**.

<u>First: What regolith properties are important to the Miners?</u>

It depends on what regolith properties are considered important to the mining method. Appendix 1 lists the asteroid surface material properties of greatest interest to the miners, which was generated during an asteroid regolith simulant workshop by Deep Space Industries and UCF in 2015 (Metzger et al., 2016). One also needs to understand how this regolith with its properties responds to the envisage action, i.e. 'touching' the asteroid. This is where our understanding of granular material dynamics in low-gravity is strongly lacking **[SKG 3]** and needs more experiments and more missions like AIDA, Hayabusa 2 and OSIRIS-REx before being sure that  we have a robust understanding and we can seriously start some systematic material extraction / utilization programs. See the additional discussion on granular materials behavior in the **Answer to Question 24**.

**11. What observable phenomenon can help constrain the potential presence of resources from ballistic experiments such a Hayabusa-II's SCI (Small Carry-on Impactor) experiment?**

Short Answer. Green, (2016)

- Enhanced features in IR spectra of dust in plume. Released volatiles hard to detect. Although we can see the plume, we cannot easily see the gases in the plume.

- Crater interior may reveal unweathered material with more diagnostic spectra. If we make a crater, will that crater be deep enough to see the volatiles?

More detailed answer.

It depends on which resource you're interested in. The ejecta pattern (this applies also to the AIDA project) will depend on the material properties. If there are volatiles, this will be obvious in the ejecta dynamics.

The Deep Impact mission was a good example for this. If we can observe the impact outcome and the plume with a camera and spectrometers, then we can then have some good information. Looking inside the crater will reveal the subsurface properties, which will also provide interesting information. If the crater fills up with fall-back ejecta, which was also a Deep Impact result, then there is also an outcrop created which shows less space-weathered material. This would probably help remote sensing by getting spectra closer to the interior of lab samples or meteorites.

Analogously the LCROSS impact on the Moon threw up an ejecta sheet and the shepherding spacecraft looked through it using visible and near infrared spectrometers to identify the volatiles in the ejecta.

> **12. How can the water absorption feature at 3.1µm be best used as an indicator of hydrated minerals on carbonaceous asteroids? What additional measurement would further increase the quality or fidelity of the measurement? What additional measurement would further increase the quality or fidelity of the measurement?**

The difficulty with the L-band, 3.1 micron diagnostic, is that 1) observations from the ground face atmospheric interference, where the spectrum is cut-off shortward of 2.9 µm. 2) We can only observe bright targets, so that rules out *realistic* space mission targets and, 3) there is a thermal emission contamination problem. So that when the observation is in the vicinity of the Earth, the temperatures are masked by the thermal emission.  To address this diagnostic properly, we need  observational capacities such as the following.

1) Extraordinary dry / high altitude sky conditions. Example: the NASA Infrared Telescope Facility (IRTF) at Mauna Kea, Hawaii  with SpeX (ref: SpEX). SpeX is an 0.7-5.3 micron medium-resolution spectrograph and imager.

2) to study the 3.1 µm feature under space-like conditions in the lab using samples that have their adsorbed water removed.  The structure of band indicates nature of material. In the next figure, we see laboratory simulations under realistic conditions, which can constrain the nature and, the degree of aqueous alteration in outer Main Belt asteroids (Takir and Emery, 2012). Studies like this are needed to calibrate the feature and estimate if there is water and how much.

3 µm feature:

- Strong absorptions in 3-µm region from interesting species (Rivkin, 2016)
    - OH ~2.7 µm
    - $H_2O$ ~2.9-3.0 µm
    - CH ~3.3-3.4 µm
    - $CO_2$, $CH_4$, $NH_3$,  carbonates…

Four types

(Green, 2016 and S. Fornasier - adapted from Takir & Emery, 2012) See also Duffard, 2016

- Sharp group (or Pallas types) exhibits a characteristically sharp 3-µm feature, reflectance decreasing with decreasing wavelength into the 2.5-2.85 µm spectral region, attributed to OH-stretching in hydrated minerals (e.g., phyllosilicates) (Rivkin et al., 2002). The majority of asteroids in this group are concentrated in the 2.5 < a < 3.3 AU region.

- Rounded group (or Themis types), exhibits a rounded 3-μm band (reflectance increases with decreasing wavelength shortward of ~3.07 μm), attributed to H2O ice (e.g., Rivkin and Emery, 2010). Asteroids in this group are located in 3.4 < a < 4.0 AU region.

- 1 Ceres-like group, is located in the 2.5-3.3 AU region and characterized by a narrow 3-μm band center at ~3.05 μm superposed on a much wider absorption from 2.8 to 3.7 μm that is consistent with brucite (Milliken and Rivkin, 2009).

- 52 Europa-like group (grouped into the Themis types by Rivkin et al., 2002)
exhibits a 3-μm band centered around 3.15 μm with longer wavelength band minimum and steeper rise on the long-wavelength edge of the absorption.

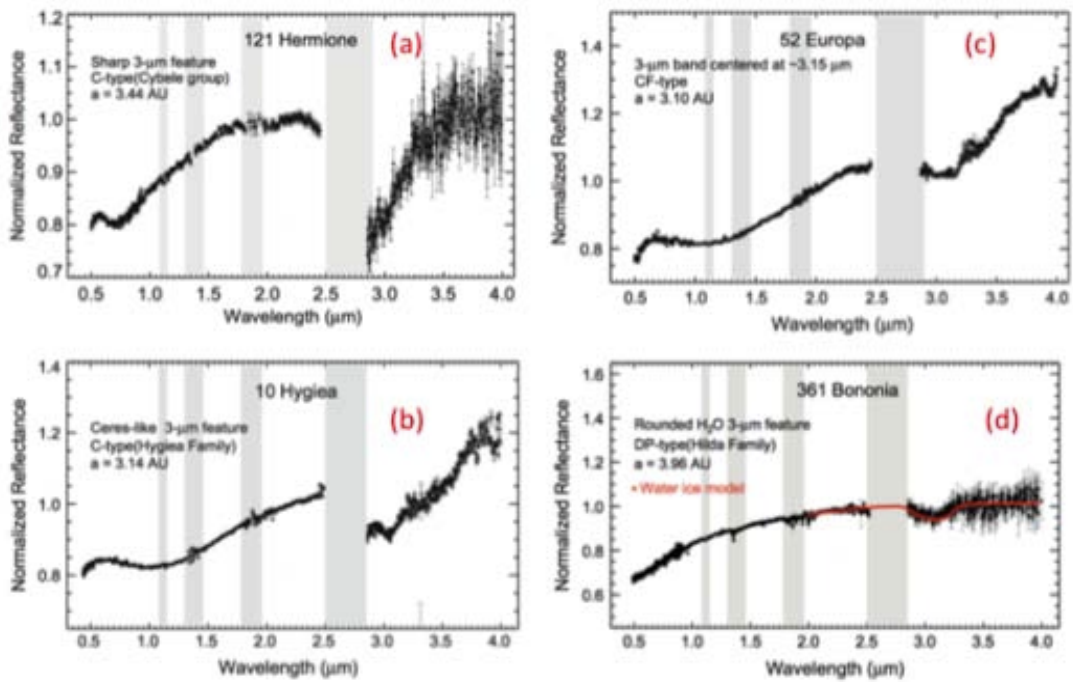

Figure 14  a) 'Sharp' Phyllosilicates or other hydrated minerals <3.0 μm b) 'Ceres-like'
2 components 3.07 μm c) 'Europa-like' No carrier identified 3.15 μm d) 'Rounded' Water ice 3.07 μm. (Green, 2016 and S. Fornasier - adapted from Takir & Emery, 2012)

2) Observations *outside/above* the thickest part of the Earth's atmosphere, such as with:

a) Stratospheric Balloon flights.

- The GHAPS platform, which is a 1m balloon-borne observatory dedicated to planetary science, may be able to access the L band during a thermally optimized ~100 day flight.

- According to P. Maier's estimates (Maier, 2016), a 0.5-m balloon-borne observatory should be able to take useful filter images of the 3 μm-region of objects down to roughly Vmag = 17 to 17.5 with reasonable observations, a 1m telescope thus should have the potential to reach around Vmag = 18.5 to 19. While the background brightness at balloon altitudes is considerably better than from the ground, it is still notable. So the performance of a 0.5 m space telescope (with freedom to point) should be above these limits.

b) Space-based.

- The size of mirror for a space-based mission needs further study. While some think that greater than a 1m mirror is necessary (Elvis), others (Rivkin) suggest that, once one is outside of the Earth's atmosphere, the measurement needn't necessarily require a mirror size any larger than what is needed to acquire a spectrum at other wavelengths. If a space mission were dedicated to getting this information, then a set of filters designed for it could probably make use of a smaller than 1 meter telescope.

- The Castaway mission (ref Snodgrass, 2016) is one possibility. This is a proposed small mission for an ESA M5 opportunity, launch ~2029.
    - Three key measurements: Spectrograph, Imager, Thermal camera
    - 0.3-5 μm survey of >10,000 asteroids of all sizes, types.
    - Spatially resolved imaging & spectroscopy in close flybys of >10 asteroids of a variety of types. Thermal IR camera for flybys
    - Discover and characterise very small asteroids (1-10m diameter) with star-tracker type cameras
    - Spacecraft is a small space telescope (50cm) that loops through the asteroid belt
    - Main telescope (50cm) with NIR spectrograph + vis. imager

- NASA will launch JWST in October 2018, with spectroscopic capabilities at this wavelength range and an effective aperture of ~6.5 m. The Integrated Science Instrument Module (ISIM, ref: JWST 2016) includes the following instruments (ref JWST, 2016): NIRCam, NIRSpec, MIRI, FGS/NIRISS.

- The MMEGA instrument on MASCOT, which is a ~10 kg Lander (DLR+CNES) for the Japanese Hayabusa-2 mission covers this range - in situ (Jaumann et al, 2016).

<u>To Address the Fidelity of the Measurement.</u>

Other features to detect water  (Duffard, 2016, Rivkin, 2016)
- There is a ~0.7 µm band associated with phyllosilcates, however it is not diagnostic.  I.e. not every spectrum that shows a 3.1 µm, also has a 0.7 µm feature.  However, if you see this feature, then you usually also see the 3.1 µm feature too.
- Mid-IR:  ~6.25 µm O-H bending vibration;   8-12 µm Si-O details.

<u>Key Point re: Hydrated minerals in asteroid (Duffard, 2016, Q&A audio)</u>
The water is bound in the mineral, such as serpentine. To extract it will take a great deal of energy.

<u>CLUE: Hydrated minerals in the meteorites (Rivkin, 2016)</u>
- Carbonaceous chondrites (CC)
  - Low albedo
  - Hydrated CM/CI/CR groups ~50% of CC falls
  - Anhydrous CV/CO/CK groups ~40% of CC falls
- "Water" bound into minerals as hydroxyl
- We don't know how representative meteorites are **[SKG 1]**
  - Hayabusa 2, OSIRIS-Rex will help
  - Not obvious what's getting screened
- There exist non-carbonaceous low-albedo meteorites, and a hydrated ordinary chondrite

<u>Hydrated minerals in the asteroids (Rivkin, 2016, Duffard, 2016)</u>
- Detected at two* wavelength regions:
  - 0.7-µm: "Ch Asteroids", detecting iron in phyllosilicates
  - 3-µm: Detecting OH/H2O per se
  - see next Figure
- Band shapes in 3-µm region show 2-3 different types of hydrated minerals
  - Phyllosilicates, like CM/CI meteorites
  - Minerals like seen on Ceres**

o Ice frost (?), organic material

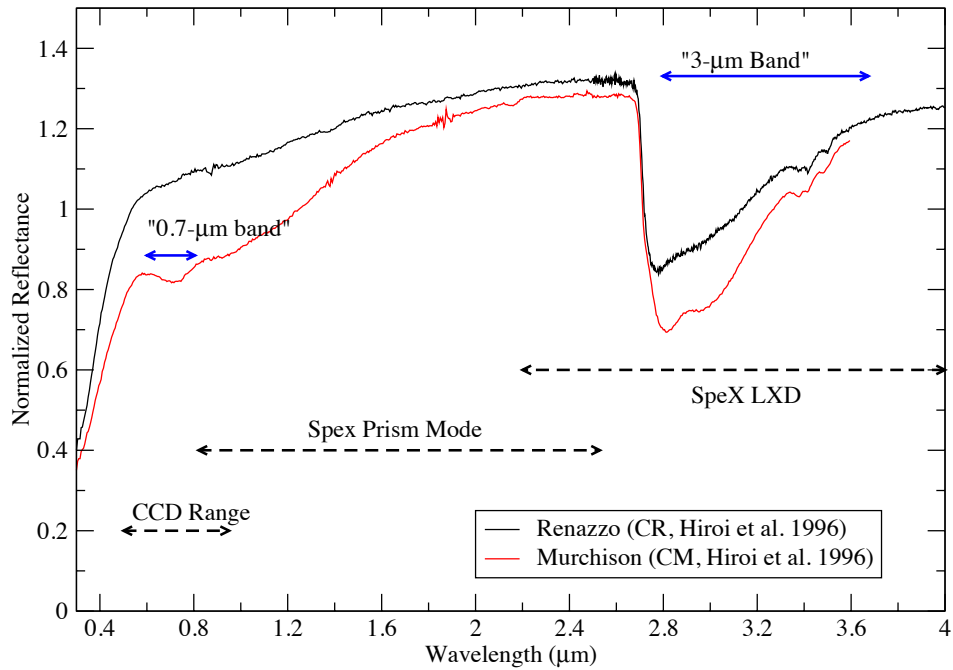

Figure 15. Hydrated minerals detected at two wavelength regions: a) 0.7-µm: "Ch Asteroids", detecting iron in phyllosilicates. b) 3-µm: Detecting OH/H2O per se

**The Ch asteroid – CM meteorite connection  Rivkin, (2016)**

- The best link between asteroids and hydrated meteorites is Ch asteroids and CM meteorites
- Both uniquely have 0.7-µm bands
- Both have similar shaped 3-µm bands indicating phyllosilicates
- Range of retrieved geochemical paramaters similar
- Similar range of albedos
- *If you want a sure thing with minimal effort, go after Ch asteroids*
- But note that they're rare among NEOs.
(This is worth further study. There are Ch types in the inner Asteroid main belt, see below and. hear the Q&A audio after Rivkin's ASIME talk)

**Hydrated Asteroids in the Solar System Rivkin, (2016)**

- Low albedo asteroids dominate asteroidal population
- Ch asteroids make up ~30-40% of C-complex asteroids
    - Pretty well mixed but more in mid belt (~40%) compared to inner or outer belt (~25-30%)
    - Larger fraction of ~100 km asteroids vs. ~30 km asteroids, fraction remains similar to ~2 km
- Very few Ch asteroids identified in NEO population
    - Surficial hydrated minerals destroyed during low-perihelion periods?

- Meteorites demonstrate hydrated minerals exist in/on NEOs

**13. How could neutron detection support prospecting activities, and what is the maximum depth at which a neutron detector could detect the presence of water?**

Short Answer: Green, (2016)

- Gamma ray/neutron spectrometer
  - Elemental abundances including C, O, Si, S ...
  - Samples to depth ~1 m. May not be deep enough to reach subsurface ice if present.
  - Low signal so spatial resolution difficult for small NEOs

One reference instrument is MSL's DAN (Dynamic Albedo of Neutrons), which can detect water ice 1 to 2 meters below the surface of Mars.

**14. What conditions would permit the presence of free water ice on an NEO (e.g., on an extinct comet), and what would be the best way to detect it remotely?**

Due to the NEO proximity to Sun, the temperature $T_{max}$ > 350 K, hence <u>no surface water</u> is possible; since it should sublimate very quickly Contrived polar cold spots are unviable.

(Rifkin, 2016):
- Ice stability very T-dependent
  - Need T ~ 110 K for surface stability
  - Need surface T ~ 140 K for shallow burial stability
  - Need surface T ~ 170 K for km-scale burial stability

- Temperature way too high for NEOs
  - 7% albedo at 1 AU: $T_{avg}$ ~ 280 K
  - 7% albedo at 1.5 AU: $T_{avg}$ ~ 230 K
  - Sub-solar T much higher

<u>The possibility of subsurface ice</u>

While asteroid subsurface ice being present for long periods (life of the Solar System) in the Main Belt at 2-3 AU is possible, such subsurface ice on NEOs wouldn't have that longevity.

In the Main Belt: depending on the thermal properties (e.g., composition, porosity, grain sizes, etc) of the dust, ice can remain for a surprisingly long period of time, potentially longer than the relatively short dynamical lifetimes of NEOs, but the NEO temperature is critical, see below. Schorghofer, (2008) demonstrated that buried ice on spherical bodies, within the top few meters of the surface, orbiting 2-3 AU from the Sun, can survive ~$10^9$ years.

For NEOs: (Rivkin, 2016)

- Ice lifetimes correspondingly short
    - Ice retreats ~330 m/My at 280 K beneath 1-km dry layer
    - Slows to 3 m/My at 230 K
    - 60 m/My at 230 K beneath 50-m layer

Note that: At 300 K, ice retreats at about 1 km per My, even under a 1 km dry layer.

So you'd need a thermal environment where the body is rarely at high temperatures, such as in a high eccentricity orbit, and it would likely need to be large enough to still retain any ice. Furthermore, the changing obliquities and low gravity of NEOs means no permanently shadowed regions near the poles.

Therefore, the combination of traits needed to retain ice for NEOs: (Rivkin, 2016)
- Very short time spent close to Sun: eccentric orbit
- Not much time spent at high temperature: dynamically young
- Sufficiently-deep insulating layer: lag deposit

In other words, you need an actual comet (and not an extinct one).

Surface impacts to expose the ice

Let's say there is more evidence that subsurface ice exists on a NEO. In principle, an impact by a micrometeorite (or something even larger) on one of these objects could briefly expose near-surface ice, which would then quickly sublimate and potentially cause an outburst of comet-like activity. It might be hard to detect in the plume, because dust scattering dominates, but an optical imaging survey looking for such outbursts (and using techniques capable of detecting extremely low levels of activity) could do it. This approach would not detect *all* NEOs with near-surface ice, but could be a way of detecting some of them.

<u>Tagish Lake meteorite pores as evidence of subsurface ice</u>

As a nugget for thought, there is some meteorite evidence of ice surviving to down to the surface of the Earth. The Tagish Lake meteorite likely had pores with at least some ice when it was collected. The Tagish samples that were curated frozen appear to retain some ice.

# II. Asteroid Surface Environment

*These questions regard the asteroid surface or near-surface, say 1 meter down for relevant detection methods. This includes: regolith, polarimetry, neutron, gamma ray spectroscopy, radar, and thermal inertia studies, space weathering, asteroid-meteorite laboratory links, electrostatic studies, shape modelers (photometry or radar could be considered 'subsurface too).*

<u>The knowledge that you need that are directly relevant to the mining aspect : i.e. composition (Green, 2016)</u>

<u>For the Surface: In-situ instruments and their measurements</u>

- Camera systems
    - Shape models; navigation;
    - Surface features, local slopes, boulders (down to ~10s of cm)
    - Multi-band imaging for compositional heterogeneity?

- Radio Science
    - Mass + Density (with size and shape model)
    - Non-uniform internal mass distribution from gravity field (Murdoch, 2016)

- Seismology (Murdoch, 2016)
    - Layering and mechanical properties of the shallow subsurface
    - Information about chemical composition and volatile content (from intrinsic scattering)
    - Depths of discontinuities, velocity and density contrasts, and subsurface layering

- VisNIR Spectrometer
    - Improved composition inference; effect of space weathering
    - lower priority since: regolith likely to be well mixed and not very diagnostic of components for mining?

- X-ray Spectrometer
  - o Improved composition inference: elemental abundances of rock-forming elements
  - o Still samples only top ~1 mm
  - o Signal/resolution issues – better deployed on surface lander

**15. Will there be opportunities to perform in situ processing experiments with asteroid regolith on future asteroid visit missions from the state space agencies, or will industry have to do this themselves?**

For NASA, this would require that the Planetary Science Division have this task among their goals, which doesn't exist at present. The work could also be in the Human Exploration Division adding it to their ARM plans, as well.

However, it depends what you mean by processing experiment. Observing how the regolith reacts to a sampling mechanism is planned!  With Hayabusa 2, OSIRIS-REx,  and a lander (Hayabusa 2, AIDA) are good opportunities to study in-situ processing.  See Delbo et al., (2016).

Even if more is needed to achieve the necessary level of understanding for mining activities. We're just beginning. Human exploration will certainly help a lot too. But if industries want to start doing more, we are more than happy, provided that we have access to the data and we can also learn from it ;-)

If ARM is not cancelled, then after the crew visit to the boulder in lunar high retrograde orbit, the boulder may be used for ISRU mining and processing experiments. NASA has issued Requests For Information (RFIs) and other statements about ARM that include consideration of asteroids as resources, so the Agency is apparently moving in that direction.

**16. Can regoliths be developed that are similar enough to the real thing that experiments would provide accurate results useful to define engineering requirements?**

Yes,  to some level of fidelity we are making it in the lab (Delbo, 2016, Metzger et al., 2016, 2016, Britt et al., 2016).  We might need to wait for the data from OSIRIS-REx and Hayabusa 2 that will tell us how well we defined the simulants for testing their mechanisms; this will give us actual data. For now, with the exception of meteorites (and their bias), and some information on Itokawa, we have no hard information on the fidelity of simulants.

Because of the many poorly known factors it is difficult to provide realistic simulant. However as one useful approach, meteorite samples and their powdered (with roughly same thermal inertia like certain measured asteroid surfaces) could be a sample type worth of using during tests. Properties of meteorite powders are a weakly explored topic however, where laboratory tests which provide useful new information that could support instrumental tests.

Case Study: Thermal cracking of meteorites / Generation of asteroid regolith

Delbo and colleagues took Murchison CM2 and an OC meteorite and temperature-cycled the meteorites and calculated what happened to the internal structure. Zooming in one of the slices with x-ray tomography, and analyzing cracks inside of the meteorite revealed the result. After one month they saw visual growth of the crack. This led to the prediction that regolith can be generated by breaking up rocks via thermal cycling (Delbo et al., 2014).

An on-going NASA-funded asteroid simulants development project recently identified 65 characteristics of asteroid materials: However, the project decided that it would be far too difficult to create a simulant with all 65 of those characteristics. Instead, the attendees of a simulants workshop decided to focus on just 12 or 13 of those characteristics. See Appendix 1.

Similarly, the lunar program identified about 30 properties and decided to simulate only two or three (particle size distribution, a measure of mechanical strength, and to a much lesser degree the mineralogy). The creators of the lunar soil simulant made no effort to replicate the important magnetic and electrostatic properties, for example, which are crucial for developing many technologies. It was decided that it would be too expensive to create a simulant that replicates most or all of the properties. This led to problems in the lunar program because simulant users often mis-used them, assuming that if it is a simulant, then it can be used in place of lunar soil for any type of activity.

The correct way to use simulants involves considering how they differ from the actual space materials, designing tests that take those differences into account, and interpreting the results in light of those difference. Simulations to extrapolate into the space environment should follow the experiment, or subsequent (more expensive) experiments that use real space materials such as lunar soil brought back in the Apollo or Lunakhod programs. If these steps are followed intelligently, then simulants can be very helpful.

<u>Considerations of making simulants sufficiently high fidelity to be useful.</u>

- First, there is little in the published literature about bulk composition of meteorites **[SKG 1]**, but there is some, and that enables us to develop simulants with all the primary mineralogy (everything more abundant than some arbitrary cutoff percentage of the bulk composition).

- Second, it is challenging, if not impossible, to replicate the organic content to high fidelity, but we can replicate it to acceptable fidelity for many purposes by using kerogen or sub-bituminous coal instead.

- Third, we do not have adequate data on particle sizing of asteroid regolith **[SKG 6]**, so regolith simulants today must be based upon rough estimations of particle size. Furthermore we have no data on subsurface particle sizing and possible layering of different particle sizes, so we cannot create simulants that can replicate that with any fidelity.

- Fourth, we can use clays that will absorb and release volatiles at a range of thermal and vacuum conditions and we can roughly match the volatile release patterns of meteorite samples, so we believe we can create that with sufficient fidelity to support engineering tests of volatile extraction.

- Fifth, we can control mineralogy to produce the correct magnetic susceptibility of asteroid materials (again based on measurements with meteorites).

- Sixth, it appears that by mixing the correct minerals, the simulant naturally produces the desired albedo, but we have not yet measured the reflection spectra to discover if the mixing of minerals naturally results in a spectrum that matches space materials.

- Seventh, the judicious choice of the phyllosilicate minerals (among the other correctly chosen minerals), with a carefully designed wetting and re-drying process, has been found adequate to create cobbles that have the desired strength properties, matching the unconfined compressive strength of actual meteorites.

There are many additional properties of asteroids that have not been replicated, but it seems that these should be adequate for most mining tests at least in the early phases of asteroid mining, as long as the users recall the limitations of simulants and do not try to use them for any purpose where they do not match the properties of space materials.

### 17. We could develop asteroid material simulants based on meteorites; how well do meteorites represent the NEO population, especially at larger (D > 10m) sizes?

If samples from Itokawa told the entire story, then reasonably good. Ordinary chondrite meteorites are good representatives of asteroids >10 m especially for S-type asteroids.

But neither meteorites nor spectral classes give the whole story **[SKG 1]**. See (Green, 2016, Duffard, 2016) We don't know the representation especially of the most fragile constituents of primitive material that may not survive the atmosphere entry. It is striking that less than 5% meteorite falls are carbonaceous chondrite meteorites, while there are so many C-types out there. Aren't we missing something? For instance the asteroid 2008 TC3 lost more than 99% of its mass in the atmosphere. Its spectrum was flat, and yet an incredible mixture of composition was found in the meteorite samples. We lack information on possible surface heterogeneities on an asteroid from ground-based observations and even more, on the inside. Clearly, meteorites are crucial, but they lack part of the story (hence the need for sample returns).

University of Central Florida and Deep Space Industries are working on that. They have a NASA SBIR to develop and produce simulants for a number of meteorite types. They currently have a CI simulant available and are developing simulants for CM, CR, CV, C2, and OC meteorites.

### 18. How well understood are the processes of space weathering, and can we tell what the original state of the surface was, based on the current state?

Space weathering is one of the processes studied in the last 20 years that shows that asteroids are not the pristine objects that we thought they were. It is a still an active area of research and but –some- corrections are being applied to understand the asteroid's original surface spectra, especially for S-type asteroids. The classic reviews of space weathering are those by Hapke (2001), Clark et al., (2002), and Chapman, (2004), which were reviewed and updated in a chapter in the Asteroids IV book by Brunetto et al., (2015).

Since the publication of the Asteroids IV chapter, among new results are useful laboratory experiments by Kuhlman et al., (2015), from Dawn space mission data to Vesta, by Blewett et al., (2016), and recent work on C-type asteroids by Kaluna et al, (2016).

From Brunetto et al., (2015) Review Asteroid IV chapter, from the observational point of view, slope trends from large surveys and in particular of young asteroid families have confirmed that solar wind is the main source of rapid ($10^4$–$10^6$ yr) weathering. The author states that the diversity and mechanisms of asteroid solar wind (SW) need to be investigated further. There are some indications that Q-type asteroids may essentially be unweathered S-type asteroids (e.g., Binzel et al. 2010; Willman et al. 2010), but the process is much less well understood for C-type asteroids. He points to a particular situation that VIS-NIR slope variations are still controversial for dark, C-type asteroids, but there's been new work since then.

Kaluna et al., (2016) present visible spectroscopic and albedo data of the 2.3 Gyr old Themis family and the <10 Myr old Beagle sub-family. The slope and albedo variations between these two families indicate C-complex asteroids become redder and darker in response to space weathering. Their observations of Themis family members confirm previously observed trends where phyllosilicate absorption features are less common among small diameter objects. Similar trends in the albedos of large (>15 km) and small (615 km) Themis members suggest these phyllosilicate feature and albedo trends result from regolith variations as a function of diameter. Observations of the Beagle asteroids show a small, but notable fraction of members with phyllosilicate features. The presence of phyllosilicates and the dynamical association of the main-belt comet 133P/Elst-Pizarro with the Beagle family imply the Beagle parent body was a heterogeneous mixture of ice and aqueously altered minerals. The apparent mineralogical differences between the Veritas family and the Themis and Beagle families highlight the importance of accounting for mineralogy when interpreting space weathering trends across the broad population of C-complex asteroids.

From Dawn space mission data to Vesta, Blewett et al., (2016), Vesta presents some open questions about V-type and S-type asteroid space weathering. The results show that as the regolith matures, it becomes darker and bluer (i.e., the 438-nm/555-nm ratio increases). This is the spectral trend predicted for addition of carbonaceous chondrite material to basalts by exogenic mixing, whereas lunar style space weathering (LSSW) should lower the reflectance but cause spectra to become redder (lower 438-nm/555-nm ratio) as npFe0 accumulates. The lack of obvious LSSW on Vesta continues to be a puzzle, because V-type and S-type asteroids are redder than powdered samples of their respective meteorite analogs **[SKG 1].** Such reddening is consistent with space weathering by solar-wind ions. While the presence of the bluish CC component in Vestas surface could partially mask the reddening effects of npFe0 accumulation, meteorite evidence suggests that npFe0 production on Vesta is indeed very low.

In 2015, Kuhlman demonstrated that the process of solar wind implantation on airless surfaces like asteroid surfaces can be reproduced in the laboratory. Kuhlman shot hydrogen atoms at solar wind speeds into tiny, polished samples of the common Solar System mineral orthopyroxene that had been placed on top of a silicon wafer. She then examined the compositional changes in the outer 20 nanometers of the implanted orthopyroxene using a scanning transmission electron microscope (STEM), and for the first time discovered the particles of iron beginning to form.

---

**19. What signatures of past water of hydrated minerals could be observed on an asteroid surface that might indicate subsurface water or hydrated minerals?**

**See Answers to Question 12 for a detailed discussion of detecting hydrated minerals in the 3.1 micron region.**

There is aqueous alteration of silicates, leading to phyllosilicates (McAdam et al., 2015). They're observable, through low-res NIR and MIR spectroscopy.

Certain meteorites are classed as Petrological Type 2 or Type 3 indicating that the minerals were altered by water on the parent bodies. These are carbonaceous. The alteration by water produced significant quantities of clay minerals, which are hydrated with hydroxyl layers in the clay mineral structure. These meteorite classes have been connected to asteroid spectral types. Therefore, simply identifying the asteroid, as belonging to one of these spectral types is the best indicator that it will contain hydrated minerals. There may be others that are hydrated that cannot be positively identified this way, but this is the best first step.

---

**20. How can the surface desiccation of carbonaceous asteroids be determined (via remote observation, in situ measurements, or theoretical models) as a function of MBA to NEO transport lifecycle?**

**See Answers to Question 12 for a detailed discussion.**

Modeling the past orbital history of each NEO would show which ones went close to the Sun for long enough to bake out any free water. **[SKG 5]** Lab experiments have shown that hydroxyl in clays can survive vacuum conditions and require fairly high temperatures to be released quickly from the clay, on the order of 400 to 500 deg C. More work is needed to determine if it can be released slowly at lower temperatures in vacuum. Current best estimate is that unless a carbonaceous asteroid was close enough to the sun to reach fairly high temperatures at or beneath the surface, then the clays at or beneath the surface should still be hydrated. This agrees with the observation of meteorites that are still hydrated when reaching Earth's surface despite being this near to the sun.

There are theoretical models for computing ice recession on main-belt asteroids (e.g., Schorghofer, 2008, 2016; Prialnik & Rosenberg, 2009). In principle, these could be combined with dynamical modeling of MBA-NEO pathways to determine the integrated total ice loss for an object over its entire migration to near-Earth space.  The remaining ice content will depend on things like the "starting" ice content of the originating main-belt asteroid and the specific thermal properties of the object (which can be estimated, but may or may not be precisely correct for any particular object) though.  This will also necessarily be a statistical assessment given the chaotic nature of the dynamical transition of a MBA to a NEO.

### 21. What proximity observations and measurements would better link remote observations to meteorite studies?

The link between asteroid taxonomic class and meteorites is a critical link. **[SKG 1]**  How near is 'proximity'?

From remote observations, we've described what is needed for linking. See Figure 2 and the presentations: (Green, 2016; Duffard, 2016, Rivkin, 2016, Delbo, 2016)  for some linking of meteorites to asteroid families.  The Science Knowledge Gap **[SKG 5]**  describes more scientific information regarding the modeling of the Main Belt source region of NEAs  for a stronger linking between the two: asteroid taxonomic class and meteorites for gaining composition information.  The discussion in **Answers to Question 1** gives a summary of the remote observational techniques needed to determine composition of the asteroid.  The following space missions will gain 'Ground truth' data between volatile asteroid taxonomic classes and meteorite types:

- Hayabusa 2 and MASCOT to C-type Ryugu,
- OSIRIS-Rex to B-type Bennu,
- AIM (with radars) for AIDA mission to Didymos?

One would presume that the asteroid is characterized 'enough' before arrival in the proximity of the asteroid. However, as described in the **Answers to Question 31** at the end have the meeting, the answer to this asteroid mining strategy: reconnaissance missions or straight to mining? is a complex answer, and dependent on the asteroid mining company's strategy.  The answer to this question also depends on the asteroid mining company's desired asteroid resource. Volatiles? Metals? See **Answers to Question 32**.

When spectroscopic studies aren't enough

If you manage to characterize the asteroid spectrascopically, it is not enough to understand the asteroid's surface regolith characteristics for a spacecraft landing **[SKG 3]**  .

For example, Eros and Itokawa have similar spectrum and albedos, if one compares a m^2 patch on each asteroid. However, the two asteroids have very different regolith properties.  See the next Figure.

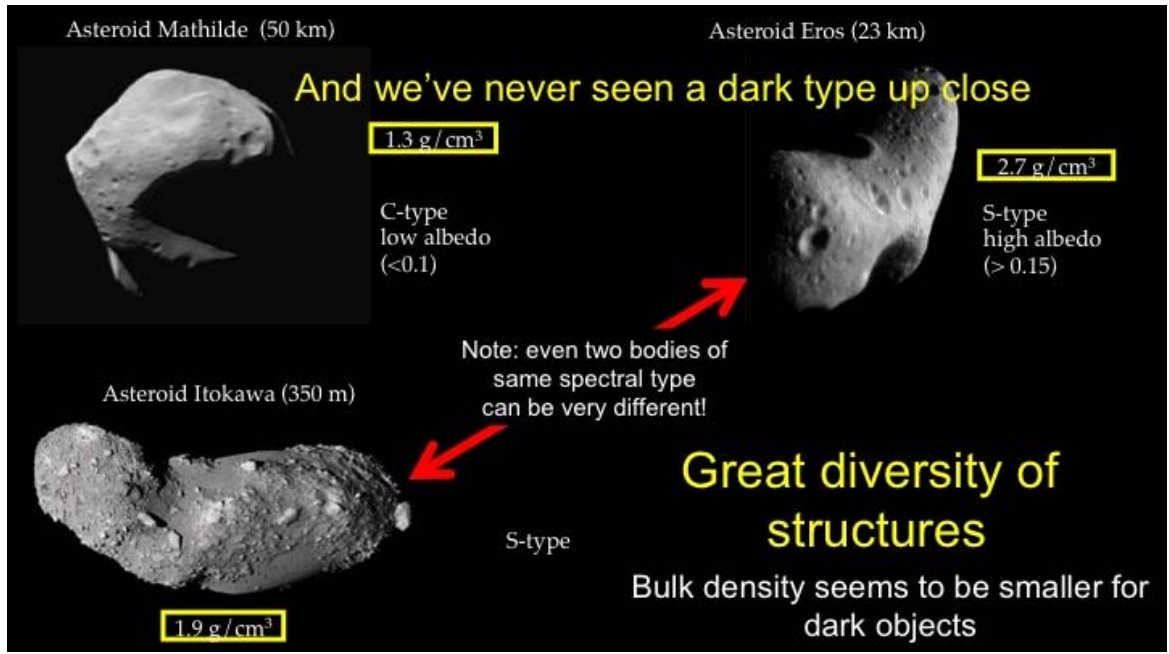

Figure 16. Comment on the scarcity of observations for C-type asteroids and why 'touching' (interacting with the regolith) is important. (Michel, 2016).

The asteroid surface regolith properties can be captured by the <u>thermal inertia parameter</u>. Thermal inertia is the resistance to temperature changes.  Eros has fine regolith size, low thermal inertia.  On the other hand, the Itokawa grain size Is the size of tens of cm, very blocky and shows a large thermal inertia. Beware of temperature effects, and compaction effects. Supporting work can be found by Gundlach and Blum, (2013)  who calculated the grain sizes for a variety of asteroids from the thermal inertia. In addition, thermal inertia measurements can also help you to <u>weigh</u> the asteroid from remote observations or proximity. See discussion in the **Questions to Answer 1**. Finally, thermal inertia measurements can also be used to identify metal rich NEAs (Drube and Harris, 2016). In that work they fit thermal infrared data with a thermal model to identify M-type asteroids while learning about its surface composition, radar albedos, spin rate, roughness of texture, and more.

## III. Asteroid Subsurface Environment

*Subsurface means asteroid interior properties. This section would include thermal modelling, rubble-pile cohesive strength studies,  collisional disruption, penetrator instruments/methods, porosity studies, quantity of water and volatiles --which ties into Nice dynamical studies  for asteroid formation location.*

> **22. Processing of mined materials will depend on composition and structure of the asteroid, and is a matter of engineering; is it necessary to develop these methods in the near future or can it be postponed until the asteroid mining industry is more mature?**

Yes, in the near future. See **Answer to Question 15**. There is already work on 'processing' if you consider interaction with the regolith surface: Real, as in space missions and microgravity experiments, and by simulation in granular experiments. Note that if ARM is not cancelled, then the boulder may be used for ISRU mining and processing experiments.

Why structure is important: You need to interact with the object to know how that object will behave and respond mechanically. Composition won't tell you how your sampling tool will interact with the surface. It is more complicated for the reconnaissance missions. These presentations discussed this topic in great detail: (Biele et al., 2016; Murdoch, 2016; Michel et al., 2016).

> **23. What are the fundamental differences between the geology of planets and asteroids?**

Planets are large enough that their gravity fields are relatively homogeneous, whereas asteroids will be smaller, with varying shapes and densities, and their gravity fields will be more varied and hard to quantify accurately before direct measurements (especially for binaries or in dense fields of asteroids). This was a problem encountered when landing probe Philae on the nucleus of comet 67P/Churyumov–Gerasimenko: even the best estimates made by experts before the landing could not expect the variability in gravity and topography.

On the surface and near-surface, at less than micro-surface accelerations, the van der Waals forces between regolith particles can exceed their weight for grains greater than mm sizes (Scheeres et al., 2010).

Asteroids are highly diverse, the occurrence and concentration of materials to mine depends on the evolution in addition the bodies' original composition. There will be differences in sub-surface deposits, with more variability expected from asteroids., but that depends partly on the parent body of the asteroid. Some asteroids came from parent bodies that were apparently larger, differentiated bodies, while others are primitive, undifferentiated bodies. Thus, some asteroids will have more in common with the geology of planets than others.

One possible great difference between larger planets' and smaller asteroids' geological evolution might be the microgravity surface - near surface processing of materials, where current knowledge has not been synthesized from the point of view what type of element concentration could happen by space weathering, by regolith turnover or segregation of different size / shape/ composition rock fragments.

Assuming similar planetary and asteroidal heating, melting, metasomatism and differentiation, internal migration and concentration of certain elements is expected. However, for for NEOs, which are fragments of larger bodies, such distributions can only be statistically estimated.

24. **How do the geological differences between planets and asteroids inform the development of asteroid mining techniques that are unique and different from terrestrial techniques? Conversely, are there terrestrial mining techniques that can be used for mining asteroids?**

Rocky planets are often differentiated, whereas asteroids will be at earlier stages of geological evolution. The presence of large variations can be expected, including the presence of gas pockets (risks of blowouts, as in marine resource extraction) or large heterogeneities (e.g. denser inclusions or boulders, making drilling more difficult or impossible). These issues are regularly faced in marine operations, and there is a large body of knowledge (and instrumentation) available to minimise risk when placing/moving assets at the surface and below. These include direct, physical intervention (e.g. Blow-Out Preventers, or BOPs, in oil and gas extraction) and remote sensing, mostly using seismo-acoustics, to detect subsurface structures and identify the ease of access to particular deposits or the risks to drilling or anchoring of structures, as well as to monitor evolution during operation (harvesting) and after (e.g. risks of subsidence). The latter can be cost-effectively monitored from space, e.g. with radar interferometry, or from the surface, with simple accelerometers or regular geodetic measurements.

For the asteroid miners, the fundamental difference between terrestrial mining and asteroid mining for tools and techniques will relate to the  underline{asteroid's low gravity.}  When one studies and numerically simulates granular materials in microgravity, there are surprising indicators for how the asteroid regolith could behave. See the examples below regarding force chains and avalalanches.

Low gravity mining techniques may include digging methods that require low reaction force or developing methods to produce the reaction force. An example of the former is a device that digs in opposite directions at the same time so forces cancel out (the method used by NASA's RASSOR low-gravity mining robot). An example of the latter is wrapping a net around the asteroid so that digging robots can grab on to produce reaction force. Also, some digging methods may be self-anchoring, pulling the robot down even as it pulls regolith up. Terrestrial surface mines usually include diggers that fill hauling trucks that travel in and out of the mine on sloped roads. Traversing in low gravity with wheeled vehicles would be extremely difficult or impossible; asteroid mining might rely upon stand-off vehicles touching the asteroid with booms rather than rolling on the surface.

<u>Presence of loose-grained materials at the asteroid surface</u>

The presence of loose-grained materials at the surface means they can be re-suspended by landers or during operations (one can think of the example of lunar regolith flying for a short time around an Apollo landing, or the longer-term suspension of dust around a helicopter close to the ground). In terrestrial mining, this is addressed by modulating the mode of extraction with the local, environmental conditions. In deep-sea mining, suspended sediments are known to move for at least hundreds of meters (e.g. Blondel, P., 2009) but there have been indications that they could be transported much further by marine currents. What will happen to loose-grained materials blown up during landing/harvesting? Will they re-depose quickly or will low gravity conditions mean they will stay up for longer?]

To this question: What will happen to loose-grained materials': It depends on how fast they are blown off; they may exceed escape velocity of the asteroid an enter into orbit around the sun, but if the delta-v is not too large then it will stay near the asteroid and may interact with it again as they orbit the sun together. Simulations have shown that the rocket plume for near-asteroid operations need have little thrust and the plume will disperse widely in vacuum before striking the regolith, therefore the drag forces on the grains are very low and the grains will not be accelerated very quickly if at all.

Thrusting very close to the asteroid, on the other hand, can produce higher velocity ejecta because the plume will not be as expanded in the vacuum and therefore the gas will have higher density producing higher drag force upon the grains. So it depends very strongly on the particulars of when the spacecraft thrusts. Also, firing thrusters directly onto the regolith can inject gas into the pore space, which then disrupts the regolith as the gas comes back out. NASA is currently funding some work on this to support ARM.

<u>Asteroid regolith can behave as if in three phases</u>

Asteroid regolith is a granular material (Murdoch, 2016). <u>Granular materials can behave as if it was a solid or as if it were a liquid and as if it were a gas.</u> Studies of granular materials studies are helpful and needed (<span style="color:red">**SKG 3)**</span>.
<u>Regolith has a memory, even in microgravity</u>
For example, inside granular materials, force chains: 'networks' build. When there are changes in the granular force network, the changes are felt over long distances in microgravity. It has a 'memory'. (Murdoch, et al., 2013; Murdoch, 2016). One perturbation at one side of the regolith could be felt on a distant location. This means that a small event e.g., a meteorite impact or spacecraft landing, on a rubble pile asteroid could easily destabilise regolith causing a granular flow. Unstable asteroid surfaces may be common.

Asteroid Seismic Activity

Asteroids are seismically active, induced from impacts. The crater-size frequency distribution can indicate how craters are being eroded for example by, e.g., seismic activity (Richardson, etal., 2005; Murdoch, N. et al, 2015 Barlow, N.G., 2015).

Asteroid regolith avalanches and slope failure

Those who have studied regolith cohesion / angles of repose on asteroids have learned that avalanches on asteroids may be more difficult to provoke than in terrestrial gravity, but when they do happen they will be larger (Kleinhaus et al., 2011, Murdoch, 2016). These researchers explain that it is because the cohesive aspect of asteroid regolith. Because the frictional force is directly related to the particles' normal force (varies as the square root of the gravity), it is harder to start the avalanche, but because there is much less friction, the avalanche will go farther.

Destabilisation of slopes and regolith migration is evident on all asteroids (Murdoch, 2016). See Figure below. Studying the slope distribution on the asteroid's surface and the regions where avalanches occurs provides information on the frictional properties of the grains and the cohesion.

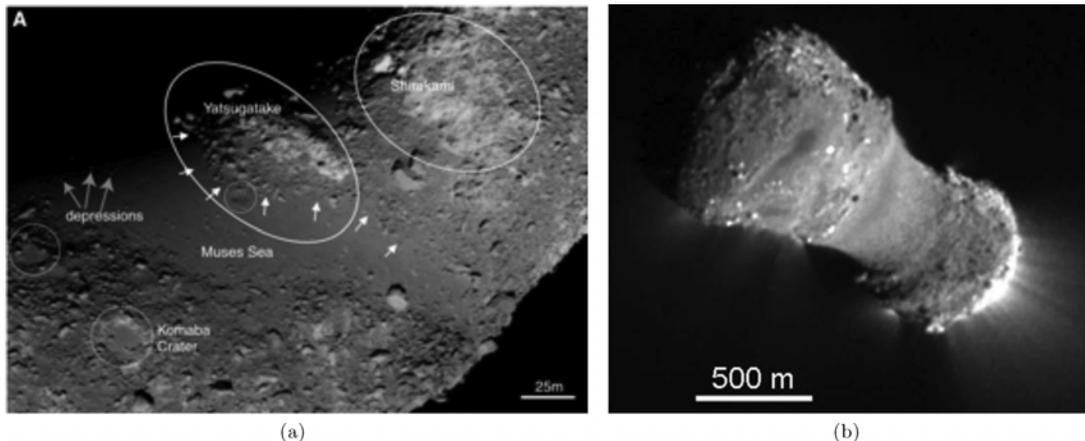

Figure 17 (a) Evidence of downslope movement on (25143) Itokawa. The small white arrows in the Muses Sea region indicate the thin, boulder-rich layer similar to landslide deposits. Image taken by the Hayabusa spacecraft; figure from Saito et al. (2006). (b) Image of comet 103P/Hartley 2 clearly showing segregation of grains on the surface. A'Hearn et al. (2011) suggest that the smooth shape of the 'waist' region connecting the two lobes might indicate material collecting in a gravitational low. Image taken by the EPOXI spacecraft; figure from A'Hearn et al. (2011). From Murdoch, et al., 2013.

**25. Is there a threshold for mechanical strength/cohesion of a rubble pile where it would be too risky to mine or interact with? If there is, how can that threshold be identified and quantified?**

Thresholds will exist: they will depend on the type of material(s) and their mode(s) of emplacement. The behaviour of these rubble piles will vary with time, as harvesting takes place, and in places identified as potential risks, it will need to be monitored closely. This can be done in situ, using tried and tested techniques like seismo-acoustics (for larger spatial scales, and with lower resolutions) and ultrasonic non-destructive testing (for very local scales, but with much higher spatial and temporal resolutions). When asteroid material has very little water content, this can also be done remotely, using microwave (radar) imaging at different frequencies (i.e. different penetration depths). See Campbell, (2002).

As explained in the **Answer to Question 24**, the destabilisation of slopes and regolith migration is evident on all asteroids (Murdoch, 2016). See the **Answer to Question 5** for ways that one might identify the strength of the asteroid's interior from its exterior manifestation on the surface.

# IV. Astrobiology/Planetary Protection

> 26. **What is the size minimum above which life might be possible (and below which life is unlikely?)**

This is a reasonable question as carbonaceous chondrite parent bodies appear to have had liquid water and moderate temperatures for millions of years, and thus represent potential sites for the prebiotic steps towards life in the first few tens of millions of years of the Solar System However, according to a study by Abramov, and Mojzsis, (2011), medium sized asteroids between 75 km and 200 km would be unlikely environments for the precursors of life. Instead, one should look to larger asteroids than 200 km and differentiated, like Ceres and the Themis family parent body for astrobiological discoveries. This is due to longer duration of heating, likely presence of long-lived liquid water oceans, and the presence of organics.

The Abramov, and Mojzsis, (2011) study modeled the thermal evolution of bodies 75, 100, 150, and 200 km in diameter: the range of sizes for C-type asteroids, and followed the heat generation through the planetesimals. The asteroid diameters were chosen to support the result of collisional evolution modeling by Morbidelli et al. (2009), which found initial planetesimal diameters from about 100 to several-100 km (asteroids are 'born big').

What follows is E. Lakdawalla's summary (Lakdawalla, 2011) for the process. All sizes of the young asteroid started with rock containing common minerals: olivine, pyroxene, water ice and pore space. The asteroids' interiors heat rapidly at first and then more slowly as the aluminum-26 decays. At an age of about a million years, the inside of the asteroid gets warm enough -- above 273 Kelvins -- for the water ice to start melting.

Once the water starts to melt, then the minerals chemically react with the water. The common minerals olivine and pyroxene react very readily, turning into a mineral called serpentine. This is an exothermic reaction, i.e. heat is released as the reaction proceeds, warming the asteroid a bit more, which melts more water, which reacts with more olivine and pyroxene, and so on; it's a positive feedback cycle.

When an asteroid's interior gets hot enough to melt water and the  runaway serpentine reaction proceeds, the center of the asteroid actually gets too hot; the 'Goldilocks' habitable zone is fairly close to the surface. As the asteroid cools off, the habitable zone migrates inward: regions closer to the center become cool enough as regions closer to the surface become too cold, freezing the ice. The cooling happens fast, so the habitable zone migrates inward at rates of around 1 to 10 millimeters per year. (It happens more slowly for larger asteroids with more initial water.) See the next Figure.

About the potential water flow while the asteroid is cooling and the habitable zone is moving. As suggested by a Bland et al. (2009) study and references therein, there may have been virtually no water flow within carbonaceous chondrite parent bodies due to extremely low permeability. This in turn suggests that even if some prebiotic chemistry were to occur, it would have remained very narrowly confined.

Bland et al. (2009) report a geometric mean pore-size of 5–50 nm observed in a primitive carbonaceous chondrite Acfer 094, and suggest that other carbonaceous chondrite parent bodies had a similar pore diameters. This presents a first-order problem for life, since the smallest known non-virus living organism, Nanoarchaeum equitans, is 400 nm in diameter (Huber et al., 2002).

So then, even though research has shown that although asteroids can be very porous, the pores are tiny (5 to 50 nanometers across) and not well-connected, so liquid water cannot migrate from pore to pore very fast, if at all. Water, and any nascent life floating in it, certainly couldn't have moved fast enough to keep up with the migration of the habitable zone. The smallest known non-virus life form from Earth is 400 nanometers across, much bigger than these pores. Even at the centers of the largest asteroids, where water would have hung around the longest, prebiotic organic chemicals wouldn't move around readily. So opportunities for reactions between naturally occurring organic molecules would be very rare.

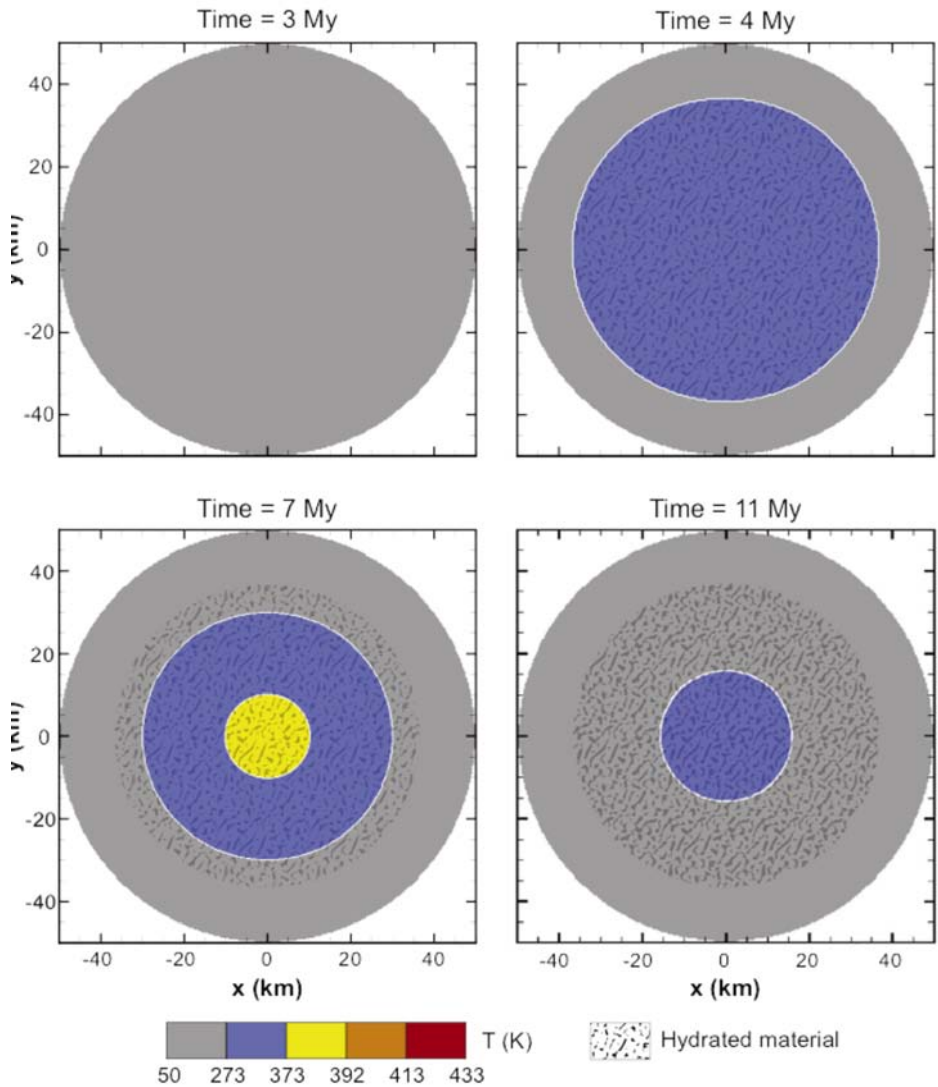

Figure 18   Temperature evolution within a model asteroid 100 km in diameter with 40% initial ice content, illustrating overall lower temperatures but longer duration of habitable temperature conditions, shown in blue. The extent of hydration is also shown. Timesteps are labeled in time after nebula collapse, at t = 3 Myr for all runs. From Abramov and  Mojzsis, 2011.

Therefore, despite the existence of habitable zones within medium-sized asteroids with diameters 75-200 km for many millions of years, the conditions for life getting started are poor.

Abramov and Mojzsis, 2011 didn't study this process for  asteroid sizes smaller than 75 km, which are those favored by the asteroid miners. However, if we assume the same conditions as in their study, then the smaller asteroids would have even more unfavorable conditions for their interiors. The cooling will be faster, so the habitable zone migration inward would be faster, with the result of a cold rock with icy, unconnected pores. After injection into near Earth space and temperatures warmer than 273 Kelvins, that ice will melt, but we still have the connected-pores problem.

Additionally, we would not see water on the surface (See Answers to Question 14). So even if we know that some extremophiles can survive in the dryness, vacuum and radiation doses of space, the conditions for getting the life started in the first place don't exist.

From an astrobiology perspective, the possible volatile and organic content of asteroids are interesting to better understand prebiotic conditions on the early Earth, but for asteroids, planetary protection is much less an issue.

We think a better question would be "what risks do asteroids hold for astronauts engaged in sampling or mining activities".  As we've described above, volatile-rich carbonaceous chondrites contain small amounts of organics that include toxic and carcinogenic compounds (Polycyclic Aromatic Hydrocarbons, PAHs) and the mineral serpentine in asbestos morphologies.  Any mining and processing plan needs to take into account these toxic hazards.

> **27. What might be the most economical and lightweight life detection method and instrument for a commercial mission to carry?**

There are many possible detection methods that could be used in a commercial mission. If we want to search biosignatures of extinct life, a good choice could be mass spectrometry. For extant life, a simple (and lightweight) method of detecting signs of active metabolism could be used (similar idea that was used in Viking mission). This would require clear distinction of abiotic chemical reactions.

# V. Other

> **28. Large- and small-scale behavior of an asteroid being interacted with is unknown. How will it respond to drilling, crushing, grinding, anchoring, rendezvousing? We don't know; so will there be opportunities on future asteroid visit missions from the state space agencies to test these behaviours?**

See the related: **Answers to Question 24**.

Hazards of Mining Asteroids (Elvis, 2016b).

- Many are "rubble piles".
    - Low densities (r ~ 1; Carry, 2012):
        - large void fraction. Ice?
    - Hard to handle,

- • Attachment what can you grab on to?
- • de-spinning
- milli-g
  - – tailings don't settle
  - – can cause problems with equipment
  - – If released will spread around orbit – hazard for Earth satellites
  - –
- Unsolved Issues in Granular physics [SKG 3 again]
  - – Speed of sound → 0. Shocks inevitable
  - – Just a few grains take the strain: unstable
  - – Harpoons likely ineffective

There is a need for investment in granular physics, including ISS experiments, to address these issues. [SKG 3 again]

The AIDA mission, if approved at the next ESA ministerial meeting, will offer some opportunities. The lander will perform some strength measurements and its surface interaction will also inform us. The impact by the DART projectile (if launched by NASA) will also allow us to get impact strength information and knowledge on how the material is excavated when we shoot on it. But so far, there's no driller or anchoring planed in any mission, as this increases the risk and complexity (therefore cost).

---

**29. Is there any mechanism for the asteroid mining industry to incorporate some of their needs into the planning of science asteroid missions by state space agencies?**

Some of us have been trying to get asteroid mining into the NASA Small Bodies Advisory Group (SBAG) Goals document (ref: SBAG goals). We are told that this year it was premature, but that continued pressure from the community can get it added. It's unclear if asteroid mining companies count as part of "the community" yet.

Perhaps it's more related to missions operations than to missions planning/design, but in Earth observation there is an ESA scheme called "ESA 3rd Party Missions": where ESA uses its multi-mission ground systems to acquire, process, archive and distribute data from other satellites . The data from these missions is distributed under specific agreements with the owners or operators of the mission, following the ESA Data Policy.

Concerning the need of asteroids exploration in a long term, on top of technological and economical roadmaps (certainly essentials), we also need a common VISION, shared between industry, scientists and agencies. Here we are talking of the future of humanity, in terms of knowledge, resources and also safety and we can all provide a critical contribution.

> **30. Are science committees and panels responsible for allocating funds for space missions aware of the enormous impact in situ resource utilization.  Which could have as an enabling technology for making structures, propellant, and life-support materials (air and water) available in large quantities in space over the long term?**

Mostly no. Some may think so privately, but it has not become real enough to them yet. There is a new paper in Space Policy:  Elvis, (2016), which describes some of these benefits.

NASA's Space Technology Mission Directorate (STMD) does fund technology development in all of these areas with the understanding that this will be revolutionary for spaceflight. The problem is that the Human Exploration and Operations Mission Directorate (HEOMD) cannot incorporate technologies into a space architecture unless those technologies are sufficiently mature. The space resource technologies have not been mature enough so they were never adopted into the prior mission architectures such as Constellation (lunar) or the Mars Design Reference Missions. HEOMD believed that, in the prior programs, it is faster and cheaper to rely on existing methods than to delay the overall program schedule while developing the immature ISRU technologies in order to possibly benefit later on.

Nevertheless, HEOMD is now including some in situ resource utilization in the architecture for Mars missions. The Mars 2020 mission will include an atmospheric capture payload to demonstrate making in situ propellants. Members of the Human spaceflight Architecture Team (HAT) are strong advocates of space mining and propellant refueling in space so they are well-positioned to make a difference. HEOMD will revise the architecture and include more reliance upon space resources ast STMD and others mature the technologies. The perception needs to be that incorporating a new technology into the Mars architecture will not slow down the schedule and prevent human landings on Mars in the mid-2030s. If propellant depots and asteroid mining can be established rapidly enough, then they can be included in the current architecture. If not, then they may be included in future Mars missions beyond the first ones in the mid 2030s.

## Questions from the Asteroid Scientists

The following section is the conference group discussion, which took place in the last two hours of the ASIME 2016 conference. The four keynote speakers sorted and prioritized the "Questions from the Asteroid Scientists" and posed these questions to the asteroid mining companies and to the audience/ participants. Here is a summary of the session speech-text transcription.

## 31. Asteroid Mining Strategy: Reconnaissance missions or straight to mining?

The companies agreed that reconnaissance missions are important to narrow down the search space. Survey missions from the ground, i.e. the "ground truth" of remote sensing can be exploited. Those preliminary ground truth missions could be government-funded science missions (Sercel). Bonin added that reconnaissance missions must also generate saleable products to make business sense. As a de-risking exercise, it might make sense spending money *before*, in a reconnaissance mission, to give sufficient confidence before is expected (Faber). You could go straight to mining if you have confidence enough (Bonin).

The cost estimation for such determination is a stochastic estimation (Sercel). The calculation is: The cost of the mining mission multiplied by the fraction of time that you fail to bring back ore versus the cost of a reconnaissance mission plus cost of waiting for the mining mission (Bonin, Sercel). If we have a mining mission to a target, and that target ends up not being viable, then the probability of that 'not viable' times the cost of that mining spacecraft is a probabilistic cost. It makes sense to do a reconnaissance mission, if the probabilistic cost of that failure of the mining mission is more than the reconnaissance mission (Sercel).

Another role for reconnaissance missions is the following (Bonin). When you calculate the probability of a mission bringing back ore, that calculation is based on scientific knowledge. That scientific knowledge could mean that you have knowledge of a number of asteroids. There could be reconnaissance missions going to a lot of asteroids to build up your knowledge.

One large caveat posed by Michel is that the ground truth between ground based and in-situ is not enough; there must be some experience of interacting with the surface. Three future planned missions will provide some of that experience. **[SKG 3]**

**Listen also to the Q&A audio after S.F. Green's presentation regarding DSI's asteroid mining company's strategy.**

## 32. Asteroid Mining Strategy: What observations can you make from the ground to avoid going to a target of which you have no interest going?

**What are minimum requirements for asteroid characterisation?**
      **- Size?, rotation? Composition (which types)?**
       **- diagnostic features from spectral features. photometry…?**

The answer to minimum requirements depends on the asteroid mining companies' desired asteroid material.

If they want regolith-like material or solid-like material, then they should be interested in low-thermal inertia or high thermal inertia material (Michel).

However, realistically, thermal inertia measurements are an impossible task (Green) for these small bodies, or these small bodies with the telescopes that we have now (Rivkin), because the small asteroids are not thermally bright enough.

Visible light curve measurements of the small asteroids are more feasible, so therefore, small body observations can show the rotation period. From the rotation period, we would know what fast spinning bodies are not useful.

Daniel Faber said that his company is interested in the material that has the highest concentration of volatiles, specifically water. It could be regolith or solid-like material if the asteroid material contains volatiles. Green pointed out that such water will be dessicated; it won't be free water ice, so it probably won't matter whether it will be in a solid or a regolith. The asteroid mining companies will want hydrated minerals. Therefore, the reconnaissance mission could provide the ground truth of volatiles being in hydrated minerals. (Green, Faber)

Green noted that less than 100-meter asteroids have little data. Near IR measurements are especially challenging because observing these small bodies from 1 AU means also removing the thermal component, which is also difficult. From the ground, Green suggested that for the small bodies, one is confined to the optical wavelengths up to 2 microns, because the 3.1 micron hydration band is hard to see. Hard, but not impossible,, according to Rivkin, who also suggests that the 0.7 micron band is a better bet and easier to observe. It is highly correlated with the 3.1 micron band. Rivkin is more optimistic about observing the 3.1 micron hydration band for the larger asteroids.

The ASIME 2016 meeting participants agreed that identifying the available low-delta-v (which are the objects with orbits similar to the Earth) targets are key. Abell suggested a map of low delta-v, low synodic period and low-albedo should be a first-cut to fine-tune the target possibilities.  **[SKG 4]**

### A short-cut to remotely characterize suitable asteroid mining targets (Idea by Delbo)

Here is an idea to narrow the parameters of studies of the low-delta-v, low inclination asteroids to specific asteroid families that produce these objects. **[SKG 5]** They can be characterized with color observations or with low-resolution spectroscopy. The dark asteroids such as those from the Polana and Eulalia asteroid families have similar compositions. Delbo suggests to choose some that are more hydrated than the other type. Then characterize a few S-types and maybe a few others to learn little better about these other families.

An elaboration on Delbo's idea (Green): First characterize the population, by observing the larger asteroids in that population, in order to get the best picture of the asteroid family of the object. So one knows for what orbital group and family is preferred. Then, second, look at the small asteroid members of that family and only choose the delta-v members that are reasonable.  **[SKG 5]**

What you don't do is spend a lot of time to look at the small objects that *don't* have the delta-v that you want. And you should be doing this at object discovery, and not 'waiting to see'.

## What instrumentation for remote observations?

Private space companies could make an investment in new/other telescope facilities (Rivkin) to improve the situation for small bodies measurements. Elvis suggested that NEOCAM could help.

## Existing Asteroid Data for Implementing a Big Data Approach

The community does not yet have an asteroids physical properties database. Nor does it have an asteroid regolith database. **[SKG 6]**

Elements exist of a physical properties database at Nice Observatory, which can be fully ready with physical properties in one-two years with funding for one-two postdocs. There are elements of a physical properties database at the MPC, but that won't be developed further. Lowell Observatory has a NASA PDART grant to implement a physical properties database to be online in 2017.

Such a database should provide the tools to make the asteroid classification, as does the SMASS web site. Observational flags should be included to note observational conditions because that influences the spectral type characteristics interpreted from the data.

Rivkin noted that there is a place for a Big Data approach for asteroids. It was used for the SLOAN data. It will be used with Gaia and with WISE. Many of the worst targets will be known this way, giving us the remainders to search for suitable asteroid mining targets.

The ASIME 2016 Meeting consensus that the asteroid mining field needs a dedicated telescope to characterize NEOs for asteroid mining. **[SKG 2]**

## Are small NEAs compositionally homogeneous as seen *remotely*?

Based on the asteroid data in hand, on a scale of 10-m square on a body, it wouldn't make any difference about spectroscopic observations during the asteroid mining orbit because it would be the same everywhere (Green). On meteorites, on a micron to cm scale, there are vast variations. When you get to boulder-sized, then we see variations too. However, Rivkin cautioned that we don't have enough asteroid data at that meter-to-ten-meter scale to give a high confidence. **[SKG 1]**

## What would be the necessary payload for in-situ observations before landing?

It depends on what you're mining for.

For a goal of volatiles: Camera and a laser altimeter (local topography).

If you're going to attach yourself to the asteroid, or try to get metal and certain organics, then you'll need different instrumentation.

Desired Instrumentation (Frank):
- xray spectrometer
- neutron spectrometer

in order to find the parameter change as a function of depth **[SKG 6].** It would give one a confidence that what one is seeing at the surface reveals what one is seeing compositionally at depth. Note that from ground-based spectroscopy: one is seeing only the surface, not the subsurface.

Suggested Instrumentation (Murdoch)
radar instrument and a seismometer instrument
(for density and texture and porosity mostly)

> **33. What are the most critical parameters for mining and for risk evaluation?  (unknown geomechanical properties, temperature variations that may be very large, consequences of low and variable gravity, electrostatics, regolith lofting, radiation, seismic activity,…  ?)**

- Regolith size distribution, and as a function of depth (Metzger) **[SKG 6]**. We need more useful published literature about the bulk composition of meteorites to help make accurate simulants **[SKG 1].** If you want robots on the surface, then you need to know cohesiveness, the regolith particle sizing, the permeability, the electrostatics environment.

- Rock mechanics **[SKG 3]** and ground-based spectra  **[SKG 2]** (Daniel Faber). Deep Space Industries'  risk need is to know where that small body sits on the line of differences between rubble pile and sand castle and a monolithic block. For the smallest asteroids, they don't need to land, they can bag them (Sercel). But there is not a way to know the interior structure, unless its spinning fast (Green). At which point, the asteroid would need to be spinning extremely fast, to be a diagnostic. Spin is generally not a useful diagnostic due to van der Waals forces between the regolith particles.

**To acquire useful Science Knowledge Gaps (SKG) from the existing literature.**

1) The NASA Human Exploration "HEOMD" and 2) NASA Small Body Advisory Group "SBAG" (ISRU) could be useful resource. Swindle says that NASA is interested in ISRU, and so an SBAG ISRU meeting is relevant. He says that NASA is not going to pay for a mining operation, but would find it valuable to conduct a study. Abell followed that 'we', meaning NASA, would be open to conducting a small study to explore SKGs for asteroid mining purposes.

**Minimoons**

NASA is looking at 2016 HO3 as a target of interest (Abell). The delta-v is too much for some people; Sercel agrees, who says that "inclinations more than a couple of degrees are really problematic."

> **34. Simulations and simulants. What role can regolith simulants play? What is the role of simulations, experimental work (drop towers, parabolic flights, plasma studies .. ) help to answer key questions about appropriate mining techniques ?**

Answer: It depends on the mining strategies. There will be a different Science Knowledge Gap (SKG) for each mining strategy.

It depends on the environment of the implementation: **risk-averse government versus entreprenurial risk takers** (Sercel). The community must find ways to embrace both methods. One is appropriate in a risk-aversive environment. And one is appropriate in an entrepreneurial environment.

Risk-averse environments will write and run the simulations that one can. To match the experiments that you can make. Then extrapolate to the planetary environment. and do precursor missions to get more data. There is never "not enough data". (Metzger)

Elaboration on Metzer's approach: To acquire data from the environment, one will need to start with experiments on Earth. But the gravity is wrong **[SKG 3]**. One will want to use simulants **[SKG 6]**, even though it is not perfect. And then keep experimenting with numerical approaches. And do all of the above to try to extrapolate from the Earth experiment to the asteroid environment. Murdoch agrees, that with their uses of numerical codes, they validate it with laboratory experiments **[SKG 3]**..

Entrepreneurial environments, where you are trying to solve a practical problem as quickly as possible, take a different approach (Sercel). You reduce the number of degrees of freedom in the problem. So that you only need to do a limited number of tests in the lab. "The best way to help the innovator is to help the innovator come up with ideas and test those ideas specifically." (Sercel)

**One potential short-cut way to know the composition from what is inside the asteroid from the asteroid's surface. (Idea by Graps)**

Use the van der Waal stickiness. We have van der Waals forces creating the cohesive forces between the blocks in a rubble pile. There should be composition evidence of those same grains leaking out of the cracks onto the surface of the asteroid. Especially the smaller is the body, which experience less exogenic processes (Green). So then you don't have to deeply investigate what is inside the asteroid, because the evidence would be on the surface. However, due to the many segregation processes, you wouldn't expect the sizes and physical characteristics to be the same. (Murdoch, Delbo)

"So in a big cake the raisins in the cake are going to be different. But in a small cookie, the raisins in the cookie are going to be the same." (Sercel)

| 35. How are the lessons we learned about mining terrestrial applicable to mining small bodies? |
|---|

On Earth. 1) exploration, 2) mining, 3) produce small particles, 4) separation. In space you would use magnetic separation, but that works only for only metals. (Terekhov)

On asteroids, the first principles are different enough, that we have to think of the problem in a fundamentally different way (Sercel).  Be careful not to be or misguided or influenced by the 1000 years of mining on the Earth (Sercel). Agreement by Terekhov.

First principles. The differences are : In space you have microgravity and copious sunlight and vacuum. These are things that you can exploit or suffer from (Sercel).

Resource assessments in space should be made as on Earth, so that the investment community can have confidence in what we are doing (D. Faber). Our resources assessments will need similar verifications to what is used on Earth. We'll need to have geologists in the room and get -them- familiar with exactly the techniques that we intend to use (Faber).

Regulations in space should allow the companies that are making those investments to be allowed to provide a return to investors (D. Faber). There's a lot of experience in the industry to get these rules right, so that is something that the asteroid mining community needs as well (D. Faber).

Resource extraction methods in space. DSI has chosen those that have the lowest technical risk, and that means that they are proven and exists terrestrially (D. Faber). We need to check that the methods work for asteroid material (D. Faber).

More generally: borrow the technology, discover the deposits, and bring it all together with the financing and scale it up to bring to the market. These are the things asteroid mining has in common with earth mining (D. Faber).

Space technology is more advanced than what we have in deep sea mining (Weiss). However, in instrumentation, there are similiarities in the space technology and subsurface sea technology.

**For mining in space with less risk. (Idea by Biele)**

Regolith has fine grains, big grains, boulders and with microgravity. Mining techniques on Earth use large machinery. Large machinery won't be useful for asteroid mining; it is very difficult in space to repair it and make it run again. We should be using very many small machinery. The machinery be of the order of the mass of the grains. So if they make mistakes, it won't mess up the process. If you have a fleet of thousands or millions of robo-ants, on the order of a cm each, the mass each of a typical piece of granular material, then if each has a simple instruction from you and can communicate with each other on the surface, then you might be able to mine that way. You can thus restrict the use of heavy machinery to just the vessel that brings back the asteroid materials.

### Timeframe and roadmap of asteroid mining activity?
(Question by Weiss)

(Answer by Sercel) The road is limited by the two needs I articulated yesterday. One is a resource model. That's a scientific need. And the second is measurement and identification of a bunch of targets **[SKG 2] [SKG 4]**. After that is technology and money. The technology is related to how fast the money flows.

(Sercel continues) So once we have the resources model to show what is the size of the accessible resources, and once we've shown that there is a future, and once we've identified a large, or sufficient number of highly accessible targets **[SKG 4],** (highly accessible means trip time in delta-v and launch dates), and that known to be water-rich, then we can start our work. That's what pacing this. So one part is technology. Another is current space operation costs, which are coming down every year,

**So, seven years. is another way to answer this question.** (Sercel)

### Topic raised by Galache: Artificial intelligence. Or how your mining equipment is going to be controlled?   (But no time to discuss)

In some cases your equipment might be sixteen light minutes away and it's not going to be a remotely operated machinery like in deep sea operations. It is not trivial to write the control systems software for this situation. There will be unexpected situations, avalanches, who knows? The control system software must be prepared for everything, literally.

# References

The most valuable asteroid science reference book for the asteroid mining companies today is the Asteroids IV book:
Michel, P., DeMeo, F.E. and Bottke, W.F. eds., 2015. Asteroids iv. University of Arizona Press.

# APPENDIX 1

The asteroid surface material properties relevant for consideration of asteroid regolith simulant at a workshop by Deep Space Industries and UCF in 2016 (Metzger et al., 2016).

1. Grain properties
      Size distribution
            Mean particle size
            Broadness of size distribution
            Coefficient of curvature (a geotechnical parameter)
            Coefficient of uniformity (a geotechnical parameter)
            Internal erodibility (a geotechnical parameter)
      Particle Shapes distribution (Morphology)
      Specific surface area
      Intra-grain porosity
2. Electrostatic properties (depends critically on the environment and is hard to replicate in a laboratory)
3. Magnetic Properties
4. Geomechanical Properties
      Fatigue
      Tensile Strength
      Compressive Strength
      Shear Strength
      Grain Hardness (hardness indexes)
      Surface friction
      Abrasivity (tool development)
      Flexural Strength-bending resistance
      Fracture properties, friability
      Impact resistance
      Rheology
      Angle of Repose
      Internal Friction
      Cohesion
      Adhesion (depends on tool material, too)
      Compressibility of regolith
      Compactibility of regolith (index test, like Proctor Compaction)
5. Physical
      Thermal properties (derived properties from mineralogy, texture, and volatile content)
            heat capacity
            conductance

thermal cracking behavior

emissivity

Bulk density of rocks

Particle density

Porosity of rocks

Surface area of rocks

Permeability of rocks

Permeability of regolith as a function of porosity/compaction

Bulk density of regolith as a function of porosity/compaction

6. Geochemical properties

Mineralogy

Organic content

C-to-H ratio (aliphatic vs aromatic)

Toxicity

Sulphur and Nitrogen content of the organic matter

Bulk chemistry (derived property of the composition)

Chemical reactivity

From surface damage

As volatile /soluble minerals

Absorptive capacity for volatiles

Isotopic ratios

Modal Composition

Siderophile elements in Iron simulants

7. Texture

Homogeneity and isotropy of texture

Chondrules

8. Volatiles

Volatiles content

Water, OH content

Organics

Sulphur compounds

Release pattern

thermal and/or vacuum release

chemisorbed, physisorbed patterns

Implanted solar wind particles (users may dope simulant if desired)

9. Optical properties

Albedo

Reflectance spectrum

Absorption

Thermal emissivity

Surface maturity (complex characteristics based on albedo and spectral shape)

10. Aerodynamic properties

Gas erodibility (rocket exhaust)
Particles' coefficient of drag

Financial support for the ASIME 2016 conference was provided by the following entities.

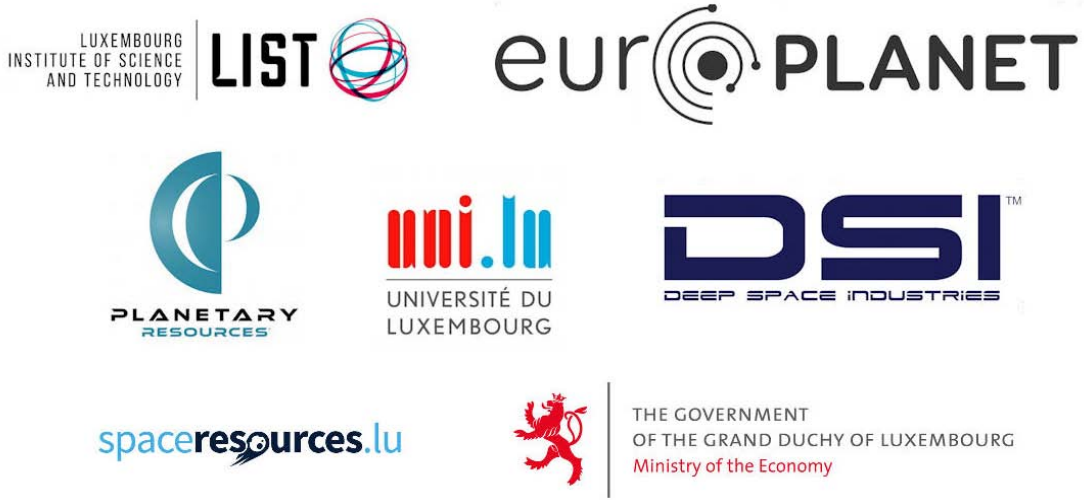